# The Attention-Directing Ability of Teams

Olga Kokshagina*
University of Sydney, Sydney, Australia
Learning Transitions, Learning Planet Institute and CY Cergy Paris University, Paris, France
olga.kokshagina@sydney.ed.au

Marc Santolini*
Sorbonne Université CNRS, ERL U1338 Inserm Paris, France
Learning Transitions, Learning Planet Institute and CY Cergy Paris University, Paris, France
marc.santolini@learningplanetinstitute.org

Christoph Riedl†
Northeastern University
c.riedl@northeastern.edu

† Corresponding author
* These authors contributed equally

**Abstract**

Why do some teams consistently mobilize collective effort and achieve superior performance while others struggle to coordinate action? We introduce Attention-Directing Ability (ADA), a latent team capability capturing how effectively members' interaction signals elicit engagement and coordinated responses from others. Extending the attention-based view, we conceptualize attention direction as an emergent coordination capability embedded in patterns of interaction rather than as a cognitive state or an outcome. Teams differ in the extent to which attention-directing signals trigger collective responses, and these differences shape how teams mobilize effort and perform. We examine ADA in a distributed innovation effort involving 2,233 participants collaborating asynchronously in 79 self-organized teams across 165 public Slack channels, generating over 30,000 messages. We model the causal responsiveness among interaction signals and derive a latent measure of ADA from teams' attention dynamics. We find that ADA strongly predicts both teams' likelihood of mobilizing engagement to sustain collective work and their performance conditional on participation. Correcting for self-selection into project submission, teams with higher ADA are more likely to submit proposals and achieve higher expert-evaluated outcomes. Causal mediation analyses show that these effects operate primarily through collective effort, indicating that attention direction functions as an upstream coordination capability. By conceptualizing attention as an emergent, measurable team capability, this study advances theories of collective attention and team coordination.

## INTRODUCTION

Teams are central to how modern organizations respond to uncertainty, innovate and solve complex problems. Decades of research on team performance have made great progress understanding how composition, shared cognition, and task structure shape outcomes (DeChurch & Mesmer-Magnus, 2010; Hackman & Morris, 1975; Ilgen et al., 2005; Lix et al., 2022; Mathieu et al., 2019). Modern teams are not stable entities executing predefined processes, but evolving systems characterized by shifting task demands, evolving contexts, and emergent states that are continuously shaped by members' behavior (Arrow et al., 2000; Cronin et al., 2011). A growing body of work views teams as complex systems in which micro-level actions accumulate, interact, and cascade into emergent patterns of coordination (Cronin et al., 2011; Humphrey & Aime, 2014; Klein & Kozlowski, 2000; Mathieu et al., 2019; Riedl et al., 2021). However, there are significant empirical challenges implementing these complex systems—in part due to insufficiently granular behavioral data and also due to the absence of modeling frameworks that connect micro-level interactions to macro-level outcomes and distinguish transient situational fluctuations from enduring collective capabilities.

In this paper, we build on the attention-based view of organizations (Ocasio, 1997; Ocasio et al., 2018; Skiadopoulou et al., 2026), which holds that organizational action depends on how attention is channeled and distributed across decision-makers. In that view, channeling is located in firms' formal structures, procedures, and communication channels and attention is allocated, largely top-down, by the context in which actors are situated. We extend this view in two ways. First, we shift the locus of attention direction from the structural channeling of attention by the organization to its emergent, peer-to-peer direction within teams: members continuously direct one another's attention through the ordinary signals of interaction, including mentions, questions, replies, pointers, rather than through formal procedure. Second, we shift the object of theory from attentional selection (what gets focused on) to attentional mobilization (how reliably a team's signals recruit uptake and response from others). Prior work on collective attention has largely operationalized attention as an amount—the number of people attending to an issue (De Domenico & Altmann, 2020; Lorenz-Spreen et al., 2019) or co-attention to a shared object (Shteynberg, 2015). We argue instead that what differentiates teams is not how much attention they hold but how efficiently attention-directing signals trigger collective response. We theorize this as Attention-Directing Ability (ADA): a latent, team-level capability capturing how efficiently a team's micro-level signals mobilize downstream engagement and goal-directed effort over time.

Critically, collective attention is inherently temporal. Attention unfolds through temporal alignment processes that shape when and how teams engage with shared problems (Riedl and Woolley 2017; Kommol et al., 2025). These temporal patterns, such as bursts, lulls, oscillations of attention, represent a critical yet understudied aspect of team functioning that shapes how teams activate and integrate resources into collective performance, which could explain why equally resourced teams diverge in outcomes. Understanding this not just as dynamical processes of attention direction but a team capability could help unlock a predictive understanding of team dynamics. Answering this question hinges on a critical empirical challenge: do the temporal patterns in team interactions reflect merely situational fluctuations or do they instead reveal a stable, latent team ability to transform resources into coordinated action? We advance two related claims: attention within teams can be directed through interaction; and the effectiveness with which teams direct attention constitutes a stable latent collective ability.

We examine how attention direction unfolded in real time and how teams' ability to direct attention triggered useful responses, recruited more effort, and resulted in performance. To examine these dynamics, we analyze a large-scale, open, and distributed innovation effort launched in response to the COVID-19 pandemic: the OpenCovid19 Initiative. Across 2,233 participants and 79 self-organized teams collaborating asynchronously in 165 public Slack channels, sending over 30,000 messages. This setting is relevant to study collective attention as many teams are now globally distributed and rely on digital technology for collaboration (Furst et al., 2004; Gibson et al., 2022). Leveraging high-resolution digital trace data, we observe attention dynamics in real time within digitally mediated, globally distributed teams—conditions that increasingly characterize organizational work. We operationalize this coordination capability of a team which we term ADA as the latent construct capturing how efficiently behavioral signals trigger collective responses, mobilize effort, shape work trajectories that influence milestones and expert-evaluated performance. The method we develop bridges a persistent gap between micro-level interaction data and macro-level team constructs. ADA helps to go beyond summarizing communication through aggregate statistics, such as message counts, response times, network density, it preserves the directional, temporal structure of who triggered what, and distills that structure into a latent team-level ability measure with a clear causal interpretation.

We have three key findings. First, we identify a consistent stable latent ability underlying teams' attention dynamics. By estimating the causal responsiveness among micro-signals in digital collaboration, such as mentions, reactions, onboarding cues, and coordination messages, we uncover a structured "attention architecture" in which some signals reliably mobilize downstream engagement while others dissipate. Second, we show that this attention architecture emerges as a latent attention-directing ability and has consequential implications for both effort and performance. Teams with higher ADA are significantly more likely to mobilize engagement and produce higher-quality outputs. We also show that ADA exhibits the temporal coherence expected of a latent team-level capability, with rank-order stability across rounds increasing as teams develop together. Third, we find that ADA operates on performance through collective effort as mechanism.

This study makes two theoretical contributions and one methodological contribution. First, we extend the attention-based view from the structural channeling of attention (Ocasio, 1997; Ocasio et al., 2018) within firms to the emergent, peer-to-peer direction of attention within teams. Whereas the attention-based view explains how organizational structures allocate attention, we show how team members mobilize one another's attention through interaction, and we recast this mobilization as a latent, team-level ability—Attention-Directing Ability. This ability identifies who triggers which response, not how much attention a team holds, as the consequential difference between teams. Second, we contribute to the growing view of teams as complex systems in which macro-level properties arise from micro-level interaction dynamics (Arrow et al. 2000; Cronin et al. 2011). ADA is a configural team property that emerges as micro-level behavioral responsiveness compounds into a stable, macro-level ability. Methodologically, we develop an approach that links this micro-level responsiveness to macro-level outcomes while preserving the directional, temporal structure of interaction—offering a template for measuring emergent team capabilities from digital trace data. These dynamics are especially consequential for distributed, asynchronous teams, where attention must be directed without co-location or formal structure.

## THEORY

### Collective attention as a team level capability

Teams operating in complex, interdependent settings face a central coordination challenge: deciding what to focus on, when, and with whom. This allows them to successfully coordinate their interactions and information integration efforts. Prior research shows that collective attention provides a foundation for integrating diverse knowledge and skills into coherent action (Kauppila & Vaara, 2026). The attention-based view locates the regulation of attention in organizational structure; we develop its team-level, emergent counterpart, in which attention is directed not by procedure but by the moment-to-moment signals members send one another. When team members align their focus temporally and substantively, they are better able to anticipate one another's actions, reduce coordination losses, and generate synergies from information exchange (Ancona & Chong, 1999; Janicik & Bartel, 2003). Collective attention shapes how teams allocate limited cognitive resources across competing issues, tasks, and social cues, shaping which problems become salient, urgent, or actionable at any given moment.

Prior work has conceptualized collective attention in multiple ways. For example, collective attention can be operationalized as the amount of attention directed toward an issue, such as the number of individuals engaging with a topic or event (De Domenico & Altmann, 2020; Lorenz-Spreen et al., 2019). Shteynberg (2015) emphasizes co-attention, in which individuals attend to the same object or event with the awareness that others are doing so as well, generating shared psychological experiences that influence motivation and judgment. However, it is not only attention to the same object that matters; teams also need to allocate attention to diverse objects to achieve coverage of important inputs and facilitate division of labor (Kozlowski, 2025; Riedl et al., 2025; Riedl & Woolley, 2017). Both the quality and the depth of attention matter to efficiently engage with all task-relevant stimuli and achieve alignment of attention across members over time and content (Woolley et al. 2023). Empirical research links collective attention to collective intelligence, learning, and team effectiveness (Chikersal et al., 2017; Tomprou et al., 2021; Woolley et al., 2023). Such shared attention emerges through social interaction, as individuals use others' cues to determine what is important and how to respond emotionally (Butterworth & Jarrett, 1991; Trevarthen & Aitken, 2001). That is, there are two interrelated processes: individuals both choose what to pay attention to, but also (attempt to) direct each other's attention to task relevant cues.

From this perspective, collective attention is enacted through interaction such as pointers, questions, interruptions, confirmations, responsiveness, and shifts in conversational focus, that signal what matters and invite coordinated engagement. Over time, these micro-level behavioral signals accumulate into patterned attentional dynamics that shape how teams prioritize work, respond to emerging issues, and sustain coordinated action.

We theorize collective attention as a foundational team-level capability that underlies and enables collaborative processes. From this perspective, collective attention is a coordination mechanism through which teams transform individual actions into collective outcomes: how effectively attention-directing signals elicit timely and sustained responses from others. We ground this view in the organizational capabilities tradition (Felin et al., 2012; Teece et al., 1997), which locates collective capabilities in micro-level actions and interactions that compose upward through routinized patterns of interdependent action (Feldman & Pentland, 2003). When directed signals successfully recruit responses, the resulting episodes reinforce sender behavior, establish responder precedents, and calibrate observer expectations—creating shared norms of responsiveness that, through repetition, stabilize into a durable team-level coordination routine. This leads us to conceptualize Attention-Directing Ability (ADA) as a latent team-level capability: the extent to which a team's micro-level signals reliably mobilize downstream engagement and coordinated activity. In complex-systems terms, ADA is a compilation emergence process: a macro-level capability arising not from members converging on a shared state but from the configuration of their signals into a responsiveness architecture that is itself the team's coordination capability. This architecture is synergistic: it mobilizes collective engagement that no individual member's signals could produce in isolation.

## HYPOTHESES DEVELOPMENT

### Attention-Directing Ability and team performance

Teams that effectively direct collective attention are better able to align engagement around shared objectives and translate available resources into coordinated action (Riedl & Woolley, 2017). Empirical studies support this view. For example, synchronization of behavioral cues, such as conversational turn-taking or affective expressions, predicts higher collective intelligence and performance (Chikersal et al., 2017; Woolley et al., 2010). Temporal "burstiness" of communication, reflected in bursts of communication around shared focal points, improves outcomes in distributed teams (Riedl and Woolley, 2017). By contrast, fragmented or weakly responsive interaction patterns, characterized by long delays, unreciprocated signals, or misaligned exchanges, impede information integration and undermine coordination (Cramton, 2001; Montoya-Weiss et al., 2001). Teams with higher ADA should therefore better translate interaction into meaningful performance outcomes.

***Hypothesis 1: Teams with higher attention-directing ability exhibit higher performance.***

### Attention-Directing Ability and collective effort

Although coordinated attention is central to team effectiveness, its influence is unlikely to act directly in all circumstances. Attention direction does not itself produce outputs; it shapes how collective effort is mobilized, sustained, and deployed. Prior research emphasizes that team performance depends critically on effort-related processes such as persistence, intensity, and follow-through, particularly in complex and uncertain tasks (Hackman, 2012). Attention-directing ability plays a central role in mobilizing such effort. By signaling priorities, highlighting relevant information, and eliciting responses, attention-directing actions help convert dispersed individual activity into sustained collective engagement and motivates members to contribute by pointing them to where their efforts are needed. When attention signals are effective, they do not just align cognition; they increase the likelihood that members contribute time, energy, and continued participation. Conversely, when signals fail to propagate, effort dissipates even when teams possess the requisite expertise and resources. This logic is consistent with process-oriented views of coordination, which emphasize that collective outcomes emerge through recursive interactions between attention, action, and effort over time (Hernes, 2014; Langley et al., 2013).

***Hypothesis 2: Teams with higher attention-directing ability will mobilize greater collective effort.***

### *Effort as a mediating mechanism*

If ADA operates by mobilizing effort, then its relationship with performance should be largely indirect. That is, teams with higher ADA should perform better not (only) because attention alignment directly produces superior outputs, but because effective attention direction sustains higher levels of collective engagement, which in turn drives performance. This reasoning aligns with prior work showing that many team capabilities exert their effects through effortful implementation rather than direct task execution (Felin et al., 2012; Teece, 2007). In the organizational capabilities' tradition, this indirectness is definitional: a capability is not the operational activity itself but the patterned ability to activate and sustain it (Helfat & Peteraf, 2003). ADA operates at precisely this level: it structures the signal-response routines through which collective effort is reliably recruited and deployed toward productive use. Therefore, ADA functions as an upstream coordination capability that shapes performance primarily by influencing how much effort teams ultimately mobilize. Performance differences should thus reflect variation in how effectively teams convert attention into sustained action.

***Hypothesis 3: Collective effort mediates the relationship between attention-directing ability and team performance.***

## RESEARCH DESIGN AND SETTING

### Context and data: Relevance of the setting to study team collaboration

Our study examines the OpenCovid19 initiative, a global open-source community launched in March 2020 in response to the COVID-19 pandemic (Figure 1). The initiative rapidly expanded from a small group of biohackers to more than 3,000 participants across 180 countries, working on projects ranging from diagnostics and protective equipment to digital and educational tools (Santolini, 2020). Coordination occurred through Slack and an online platform, enabling decentralized organization into project teams and working groups (Kokshagina, 2021). Participants contributed by self-selecting into projects aligned with their expertise and interests, while results were openly shared to encourage replication and adaptation. Figure 1a outlines the evolution of the overall activity in the OpenCovid19 Slack workspace. Figure 1b indicates how microgrant processes were organized through peer evaluation, scoring, feedback, and funding.

OpenCovid19 initiative provides a particularly valuable setting for examining the dynamics of collective attention and team performance in contemporary organizations. At first glance, a global, open-source community of biohackers responding to a pandemic might appear to be an unusual case. Yet, this context closely mirrors conditions increasingly characteristic of modern teamwork.

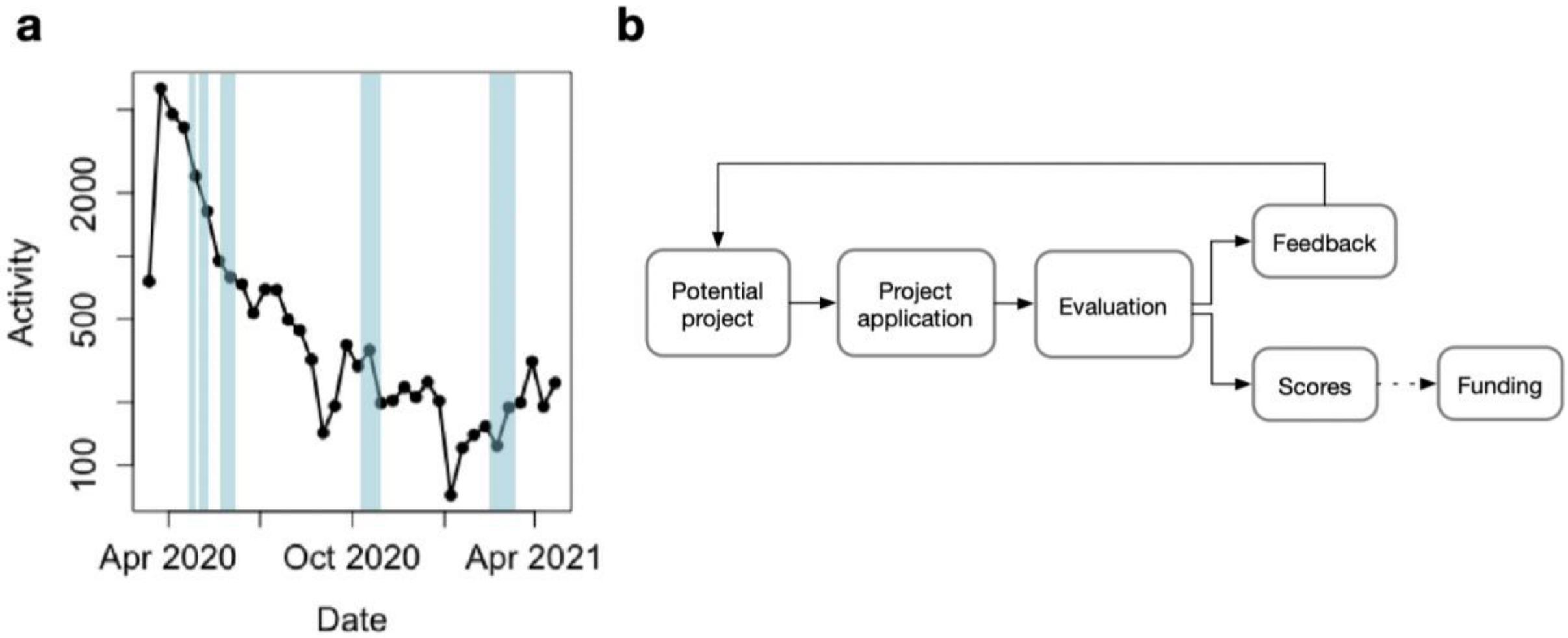

**Figure 1. Program activity and microgrant process. (a)** Temporal evolution of overall activity in the OpenCovid19 Slack workspace, measured as the number of unique Slack events per time unit. Vertical blue bands indicate the five microgrant periods during which project proposals could be submitted. **(b)** Schematic of the microgrant process, from potential project emergence through application, peer evaluation, scoring, feedback, and funding, adapted from Graham et al. (2023). This process links collaborative activity to externally evaluated outcomes and motivates the subsequent analysis of attention dynamics and team performance.

First, today's teams operate in environments where remote and hybrid work has become the norm rather than the exception. In many industries, employees spend a substantial share of their time working offsite, with hybrid arrangements combining on-site and remote activities growing rapidly (Bloom et al., 2024). Teams are often geographically dispersed, spanning multiple offices, countries, and time zones (Raghuram et al., 2019). Coordination relies heavily on digital infrastructures, including Slack, email, project management platforms, and shared calendars, that facilitate collaboration across distance but also introduce new challenges of synchronization and attention management (Yang et al., 2022).

Second, teams in traditional firms increasingly face volatile and uncertain environments, requiring them to adapt quickly to shifting priorities and external shocks. Scholars describe this as part of the "remote and hybrid era" of work, where technological change, crises, and global interdependencies amplify uncertainty and require rapid collective problem solving (Kossek et al., 2023). In this respect, OpenCovid19 offers a rare opportunity to observe how teams mobilize collective attention when all projects launch simultaneously, address a common high-stakes problem, and are evaluated by shared criteria of success.

Third, the OpenCovid19 setting captures features of open and boundary-spanning collaboration that are increasingly salient in corporate environments. Traditional organizations now routinely involve cross-functional, multicultural, and multi-organizational teams to tackle complex challenges (Bailey et al., 2022). These teams, like the OpenCovid19 project teams, must integrate diverse expertise and reconcile competing perspectives while working under time pressure and with high interdependence. By studying settings like OpenCovid19, we can gain analytical leverage on issues that are equally pressing for teams in more traditional firms: how attention is directed and coordinated in distributed work, how team members synchronize across temporal and geographic divides, and how collective effort is sustained in uncertain contexts. In short, OpenCovid19 represents a microcosm of the modern organizational team, making it a compelling setting to advance theory on collaboration and team performance in the remote and hybrid era (Blay et al., 2024; Kossek et al., 2021).

Our dataset is based on Slack communication of the OpenCovid19 initiative. In total, 2,233 participants interacted across 165 public Slack channels. We identified 79 project-related channels (either prefixed with *proj-* or mapped to a microgrant record), of which 30 could be linked to the microgrant evaluation dataset. Slack's event log includes messages and directed behavioral signals including mentions, reactions, pinned posts, and onboarding actions, providing a fine-grained view of how participants directed one another's attention over time. After restricting to project-related channels during microgrant periods, the dataset contains 13,912 unique Slack events, corresponding to 29,218 coded action instances (e.g., a message containing both a mention and a URL contributes multiple action codes). These actions are linked to grant-round outcomes, which include submission indicators and reviewer scores, across five grant rounds (Figure 1), yielding 240 team–round observations, of which 44 include an evaluated submission with an associated score.

**Dependent variable: team performance**

Our primary dependent variable captures team performance through an externally evaluated and consequential measure of collective success. The microgrant process required teams to submit project proposals that were assessed by multiple independent reviewers with relevant domain expertise using a structured community review system (Figure 1b). This review process was specifically designed for rapid and scalable resource allocation in large, open innovation communities and has been formally validated in prior work (Graham et al., 2023). Because evaluation outcomes determine whether projects receive formal recognition and financial support, they constitute a meaningful indicator of teams' ability to translate collaborative activity into externally valued outcomes.

We operationalize team performance along two related dimensions. First, we measure grant submission, indicating whether a team submitted at least one proposal during a given grant round. Submission reflects a team's capacity to mobilize collective effort, converge on a coherent project articulation, and engage with a formal evaluation process under time constraints. Second, conditional on submission, we measure evaluation score, defined as the average numeric rating between 0 and 5 assigned by reviewers to each submitted proposal. When teams submitted multiple proposals in a given round, we retain all submission-level scores and link them to the corresponding team–round observation.

This two-stage structure—participation followed by evaluation—is theoretically and empirically important (cf. Jeppesen and Lakhani 2010). Not all teams choose to submit proposals in every round, and prior work shows that repeated rounds allow teams both to advance promising early prototypes and to iteratively improve weaker proposals over time (Graham et al., 2023). At the same time, among submitting teams, evaluation scores capture qualitative differences in project clarity, feasibility, and perceived impact as assessed by independent reviewers. Together, these outcomes allow us to distinguish between teams' ability to enter the competition and their ability to perform well once they do, a distinction that is central to our analysis of attention-direction ability.

**Independent variables**

***Attention-Directing Ability***

*Operationalization*

Conventional team communication measures, such as message counts, response times, network density, capture how much teams communicate, not how effectively their signals mobilize responses. In contrast, our approach aims to shows that the heterogeneity across teams is not idiosyncratic but reflects a single underlying capability dimension.

We operationalize Attention Direction Ability (ADA) in three steps (Figure 2 for more details). First, we estimate the causal responsiveness between all ordered pairs of action types by fitting distributed-lag negative-binomial mixed-effects models to the hourly time-series panel; this yields, for each team and grant period, a set of random slopes capturing how strongly each trigger action tends to elicit downstream responses. Second, we summarize these pair-level slopes at the action level by averaging, for each trigger action, its positive random slopes across all response actions, producing a compact responsiveness profile for each team-period. Third, we assess the latent structure of these profiles using principal component analysis as a dimensionality diagnostic and extract a single dominant factor via exploratory factor analysis; the resulting factor scores constitute our measure of ADA.

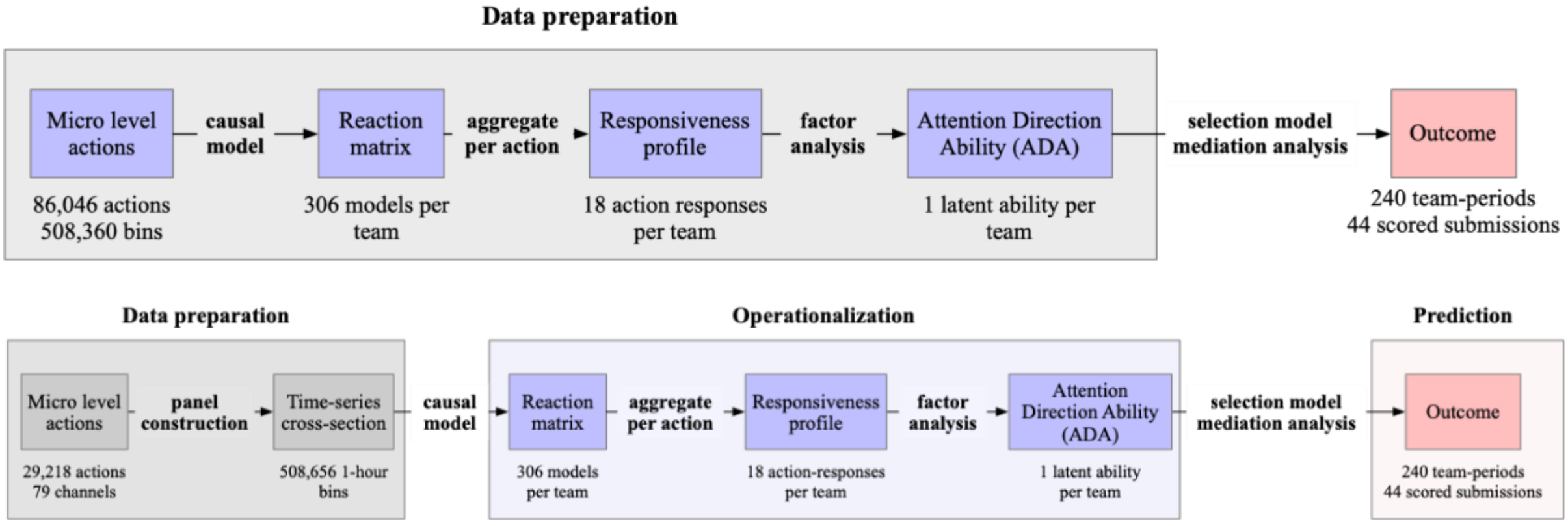

**Figure 2. Analytical pipeline from micro-level actions to performance outcomes.** Starting from high-resolution digital traces extracted from Slack interactions (29,218 actions across 79 channels), we first create a time-series–cross-section of 508,656 one-hour bins. We then estimate causal action–response relationships using distributed-lag negative-binomial mixed-effects models for all ordered pairs of action types. This step yields, for each team and grant period, a reaction matrix capturing how strongly activity in one action predicts subsequent activity in another (306 models per team–period, corresponding to all ordered pairs drawn from 18 action types). Next, these reaction matrices are aggregated at the action level to produce a responsiveness profile for each team, summarizing how effective each class of behavioral signals is at eliciting downstream engagement across responses. We then apply factor analysis to these team-level responsiveness profiles to extract a single dominant latent dimension, which we interpret as ADA, a team-level capability reflecting how efficiently micro-level signals mobilize collective engagement. Finally, ADA is used as the key explanatory variable in selection-corrected outcome models and causal mediation analyses linking attention direction to collective effort and performance.

By exploiting within-team, within-period variation at hourly resolution (while absorbing platform-wide activity shocks, calendar regularities, and persistent differences in team baseline activity) the pipeline isolates local interaction dynamics from the confounds that plague observational studies of team communication. What remains in the random slopes is not 'how much does this team talk' but 'how effectively do this team's signals mobilize follow-on engagement. This distinction is illustrated in Figure 3, which contrasts a low-ADA team whose signals dissipate without triggering coordinated responses with a high-ADA team whose mentions, reactions, and onboarding cues reliably cascade into sustained collective activity. Crucially, the method makes no a priori assumptions about which actions matter for attention direction. Instead, it estimates mobilization effects for all plausible action pairs and lets the factor model discover which contribute to a shared latent dimension. The resulting measure is empirically grounded rather than theoretically imposed: the data determine both the weights and the dimensionality.

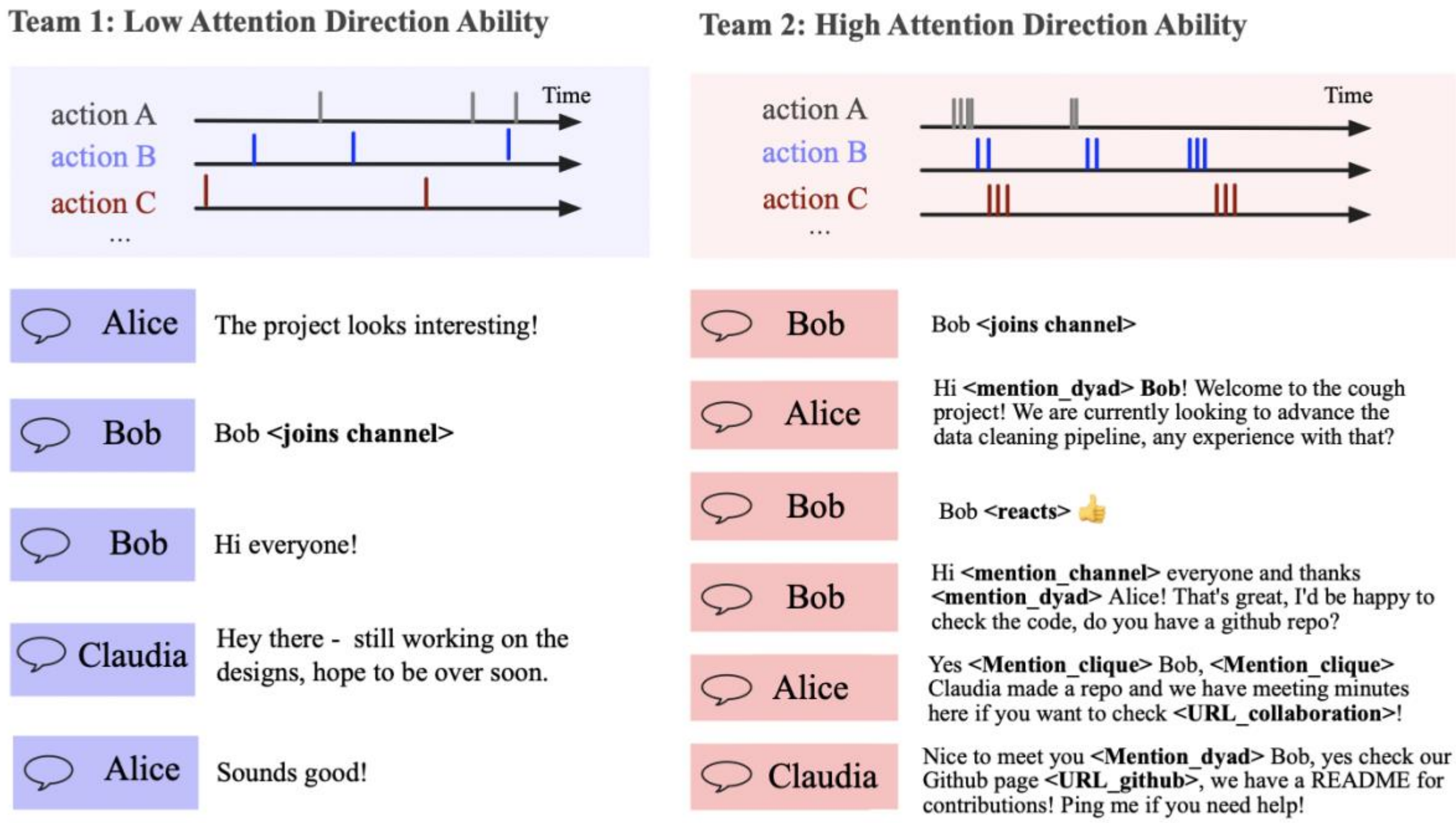

**Figure 3. Illustrative interaction patterns underlying low versus high ADA.** This figure contrasts two stylized team interaction patterns to illustrate how Attention-Directing Ability manifests in practice. (Left) Low-ADA configuration: attention-directing actions (e.g., first direct interactions, dyadic mentions, information sharing) fail to generate engagement and lead only to delayed or fragmented downstream responses. (Right) High-ADA configuration: signals reliably trigger timely replies, reactions, and further coordination, producing short-term cascades of collective engagement.

*Step 1: Estimating attention-triggering relationships between actions*

To estimate how behavioral signals trigger subsequent activity, we use a series of distributed-lag mixed-effects models (Gelman & Hill, 2007). Action types are defined by the authors based on communicative acts on the Slack messaging system given by its native taxonomy[1] and extended through additional functional recording to capture attention-directing behaviors. While some actions correspond directly to Slack events (e.g., messages, replies, reactions), others are researcher-defined abstractions that decompose events into analytically meaningful components, such as dyadic or clique mentions, onboarding indicators, and distinct categories of shared URLs (e.g., scientific, coding, or coordination resources). These action types reflect different ways in which participants may direct, signal, or respond to attention within a team. Action types are defined based on Slack's event taxonomy and capture distinct attention-directing behaviors, including conversational, coordination, onboarding, and information-sharing actions; full definitions are provided in *Supplementary Information*.

[1] Source: https://slack.com/help/articles/360002063088-Understand-your-actions-in-Slack accessed 2026-03-23, Accessed Pctober 5, 2025

Using this action set, we estimate, for all ordered pairs of action types ($A_t \rightarrow B_{t+1}$), how the count of action $A$ in time bin $t$ predicts the count of action $B$ in time bin $t+1$. Time bins are set to one hour, a resolution motivated by the empirical distribution of inter-event times in the data: Slack communication in the OpenCovid19 workspace is highly bursty, with a median inter-event time of approximately 2.8 minutes but a mean of 1.5 hours, ensuring that most causally connected interactions fall within the same window while generating a sufficient number of non-empty bins for stable estimation (Supplementary Figure S2). Examples include whether mentions trigger replies, whether onboarding events trigger messaging, and whether information-sharing actions elicit further coordination. At this stage, no action pair is assumed to be attention-directing *a priori*. Instead, for each ordered pair, we fit a separate distributed-lag model testing whether the count of action $A_t$ during time bin $t$ predicts the count of action $B_{t+1}$during time bin $t+1$. With $K$ action types, this procedure yields $K$ x $(K - 1)$ models (306 models in the baseline specification), each estimated using an identical structure. To illustrate the kinds of interaction dynamics these models capture, Supplementary Table S3 presents illustrative action–reaction sequences examples drawn from high-performing teams (grant score > 4.5). We focus on interactions within the core interpersonal coordination network (i.e., dyadic mentions, first direct interactions, replies, and messages) where attention is most observable. The examples illustrate how attention-directing signals trigger downstream engagement, recruit expertise, and generate coordinated action.

Each distributed-lag model includes the same fixed and random effects. Fixed effects account for platform-wide activity shocks and temporal regularities, including (i) total Slack wide activity in the same time window, (ii) the number of active users' platform-wide, (iii) day-of-week indicators, and (iv) calendar-month indicators (see section *Control variables*). Random intercepts are specified at both the channel and *channel* × *grant-period levels* to capture persistent differences in baseline activity intensity across teams and project phases.

Each model includes a random slope on the lagged predictor (action $A$) that varies by *channel* × *grant-period*. This random slope captures how strongly a given team, during a specific phase of its project, tends to respond to a particular attention cue. A larger positive random slope indicates that when this team emits action $A$ (e.g., a dyadic mention), it reliably elicits more follow-on activity of type $B$ in the next time window, whereas a slope near zero indicates that the same signal tends to dissipate without mobilizing responses. This specification allows the lagged coefficients to be interpreted as causal impulse responses, identified by within-team, within-period variation in local interaction dynamics, rather than by global fluctuations in platform activity or calendar-driven cycles (Wooldridge, 2010).

All models are estimated in R using the *lme4* package (Bates et al., 2015) using a negative-binomial specification to match the over-dispersed counts with many zero observations of the time-binned and grouped panel structure.

*Step 2: Constructing action-level responsiveness profiles*

During the first step, the random slopes represented heterogeneity in teams' responsiveness to specific signals; they do not yet constitute a latent ability. In the second step, we turn the estimated random slopes into an action-level responsiveness profile for each *channel* × *grant-period*. Specifically, for each trigger action $A$ within a given team-period, we aggregate information across all response actions $B$ by computing the mean of the positive random slopes from the corresponding *(A → B)* models. Averaging only positive slopes ensures the resulting profile reflects genuine mobilization rather than conflating attention-triggering effects with suppression or null associations. This average captures how effective action *A* is, on average, at mobilizing any downstream activity within that team-period, while ignoring negative or null associations that do not reflect attention-triggering dynamics.

The result is a compact representation of each team-period's responsiveness profile: a vector indexed by trigger actions, where higher values indicate action types that, for that team and phase, tend to elicit stronger follow-on engagement across multiple behavioral channels. This step reduces the high-dimensional space of pairwise interactions into an interpretable action-level summary while preserving heterogeneity across teams and project phases.

*Step 3: Extracting a latent attention-direction factor*

In the third step, we assemble the resulting *ChannelPeriod* × *Action* matrix and assess its latent structure. As shown in the *Results* section, a principal component analysis (PCA) reveals a single dominant dimension that clearly exceeds a shuffled null benchmark, indicating that teams' responsiveness profiles share a common underlying structure rather than reflecting unrelated pairwise effects. This shared structure constitutes what we term *synergistic responsiveness*: the component of collective engagement mobilization that arises from the combination of directed signals and responses and is irreducible to any single signal type's individual effect (Rosas et al., 2020). The latent factor thus captures the higher-order dimension along which the action–response couplings cohere—the coordination structure that makes the team's ensemble of signals more than the sum of its parts. Guided by this diagnostic, we estimate a one-factor exploratory factor analysis (EFA) model on the same matrix.

The resulting factor scores define Attention-Directing Ability (ADA): a latent team-period–level capability capturing how efficiently a team's ensemble of micro-level signals translates into subsequent collective engagement. For interpretability and comparability across analyses, ADA is linearly rescaled to lie between 0 and 1, with higher values indicating teams whose attention cues consistently elicit downstream activity across multiple behavioral channels.

## Control variables

The analyses include a set of control variables introduced to address identifiable sources of heterogeneity in activity levels and temporal structure that could otherwise confound (i) the estimation of action-to-action responsiveness used to construct ADA and (ii) the downstream analyses linking ADA to communication effort, submission behavior, and evaluation outcomes. The inclusion and specification of these controls follow directly from known properties of digital communication data and from the institutional features of the OpenCovid19 process.

*Interaction-level controls*

At the interaction level, i.e., in the distributed-lag mixed-effects models used to estimate action-to-action responsiveness, we control for total platform activity and the number of active users in each time bin. These variables account for exogenous fluctuations in overall participation that affect action counts across all channels simultaneously, independent of any given team's internal coordination dynamics. Both quantities display highly skewed, heavy-tailed distributions characteristic of online collaboration systems. To reduce the influence of extreme values and stabilize estimation, we apply a $log(x+1)$ transformation and standardize these variables prior to model fitting. Including these controls ensures that estimated attention-triggering effects reflect relative responsiveness rather than background surges in platform-wide engagement.

We also include day-of-week and calendar month indicators to adjust for systematic temporal rhythms in communication behavior. Prior work on digital work and online communities documented strong weekly cycles and seasonal variation in activity levels. Modeling these effects explicitly allows us to separate short-term interaction dynamics from predictable calendar-driven fluctuations (Wooldridge, 2010).

*Team–round controls*

At the team–round level, i.e., in analyses linking Attention Direction Ability (ADA) to submission behavior, evaluation scores, and mediation through communication effort, temporal controls are adapted to the structure of each model. In the mediation analyses, we include grant-round fixed effects to account for systematic differences across microgrant rounds, such as variation in participation intensity, reviewer load, or evaluation conditions. This specification ensures that mediation estimates are identified from within-round variation in teams' attention dynamics and effort. In the selection using Heckman models (Greene 2012; Heckman 1979), we instead control for the number of days since channel creation, which captures teams' temporal proximity to submission deadlines and their opportunity to mobilize activity. Because days since creation are highly correlated with grant rounds, including both would risk overadjustment. We therefore use days since creation as the primary temporal control and as the exclusion restriction in the selection equation.

## Analytical strategy for outcome analysis

Having constructed ADA as a latent team-level capability, we next examine whether and how this ability predicts downstream performance outcomes in the OpenCovid19 microgrant process. Two features of the empirical setting shape our analytical strategy. First, grant evaluation scores are only observed for teams that choose to submit proposals, implying that any analysis of performance conditional on submission is subject to non-random sample selection. Second, our theory posits that ADA operates as an upstream coordination capability that mobilizes collective effort, suggesting that its effects on performance should be largely indirect. To address these challenges, our outcome analysis proceeds in two complementary steps.

*Selection-corrected performance analysis*

Because teams self-select into proposal submission, observed evaluation scores are potentially biased by unobserved factors that jointly affect the likelihood of submitting and expected performance. Any regression that conditions on submission without correcting for this process risks omitted-variable bias. Following established approaches to innovation contests and distributed problem solving (e.g., Jeppesen & Lakhani, 2010), we therefore estimate a two-stage Heckman selection model (Greene 2012; Heckman 1979). The selection equation models the probability that a team submits a proposal in a given grant round, while the outcome equation models the evaluation score conditional on submission. Identification relies on an exclusion restriction based on project timing: the number of days between channel creation and the submission deadline, which plausibly affects teams' opportunity and readiness to apply but does not influence reviewers' assessments of proposal quality. This approach allows us to estimate the association between ADA and performance while correcting for endogenous selection into the observed sample of submissions. Models are estimated in R using the *sampleSelection* package (Toomet & Henningsen, 2008)

*Causal mediation analysis*

We examine the mechanism through which ADA affects both participation and performance. Our theory predicts that ADA mobilizes sustained collective effort, which then drives performance. To test this mechanism, we conduct causal mediation analyses (Imai, Keele, & Tingley, 2010), using effort (measured as total message volume) as the mediator. This framework estimates indirect effects by comparing model-based counterfactual outcomes under alternative values of the mediator while holding ADA constant, allowing us to distinguish ADA's indirect effects operating through effort from any residual direct effects. Models are estimated in R using the *mediation* package (Imai, Keele, Tingley, et al., 2010) with bootstrap confidence intervals (*B=1,000*).

To assess the robustness of the mediation estimates to potential unmeasured confounding, we additionally conduct a sensitivity analysis using the *medsens* function from the *mediation* package. This procedure evaluates the sensitivity of the average causal mediation effect to violations of the sequential ignorability assumption by varying ρ, the hypothetical correlation between the error terms of the mediator and outcome equations induced by an unmeasured confounder. For each value of ρ, the analysis computes the Average causal mediation effect (ACME) that would obtain if such a confounder were present, along with bootstrap confidence intervals. The value of ρ at which the ACME crosses zero and the corresponding proportion of residual variance in both equations that such a confounder would need to explain provides a quantitative benchmark for evaluating the plausibility of confounding as an alternative explanation for the mediated effects.

## RESULTS

### Descriptive statistics

We estimated 306 negative-binomial mixed-effects lag models, one for each ordered pair of action types *(A→B)*, using a common analytic panel of 508,656 one-hour bins spanning 240 *Channel×Grant-period* groups and our set of controls to absorb platform-level shocks. Model fit varied strongly across interaction pairs (*log-likelihood −21,284 to −163; AIC range = 42,242*), indicating that some behavioral transitions are highly structured while others are comparatively stochastic at the one-hour resolution. Predictive alignment between fitted conditional means and observed counts was typically modest but nontrivial (median *cor(fit,obs) = 0.160; IQR = 0.130–0.281; range = 0.055–0.706*). This pattern is expected in bursty, heavy-tailed communication streams where most bins are zero and where the models use only a one-step lag plus global temporal controls: the goal is not out-of-sample prediction accuracy but consistent estimation of responsiveness parameters that distinguish interaction dynamics across teams and project phases. Fixed-effect estimates were summarized distributionally to characterize general tendencies rather than to test individual hypotheses: total platform activity showed the strongest association with downstream action counts, with a median Incidence Rate Ratio (IRR), i.e., the multiplicative change in expected count per unit increase in the predictor, of *3.14 (IQR [2.39, 4.22]),* whereas the number of active users and lagged within-channel actions were smaller on average *(median IRR = 1.02, IQR [0.78, 1.60] and 1.01, IQR [0.99, 1.03], respectively)*. These diagnostics motivate the subsequent analysis: we treat the collection of channel-period specific random slopes as a high-dimensional

"attention-triggering profile," and we test whether its shared variance collapses onto a low-dimensional latent capability that predicts effort and performance.

Table 1 presents descriptive statistics and zero-order correlations among the team-level variables. ADA exhibits substantial variation across team–round observations *(N = 240; M = 0.10, SD = 0.19*), indicating meaningful heterogeneity in teams' propensity to elicit downstream activity following attention-directing actions. As background context, project channels also varied widely in the number of active participants *(median = 11 participants, IQR = 3–18; range = 1–113)*, reflecting the open and self-organized nature of collaboration. ADA is strongly correlated with message volume *(r = 0.76, p < .001)*, suggesting that teams scoring higher on ADA also tend to be more communicatively active. ADA is positively associated with both the likelihood of submitting a proposal *(r = 0.24, p < .001)* and, among submitting teams, with evaluation scores (r = 0.38, p < .05). Message volume is likewise positively correlated with both submission (r = 0.30, p < .001) and score (r = 0.53, p < .001). Grant round shows negative correlations with ADA and message volume, consistent with declining activity across successive rounds. These correlations are descriptive in nature and do not imply causal relationships; formal tests of the proposed mechanisms are reported in the subsequent multivariate, selection-corrected, and mediation analyses.

| Variable | N | M | SD | Min | Max | 1 | 2 | 3 | 4 | 5 |
|---|---|---|---|---|---|---|---|---|---|---|
| **Attention Direction Ability** | 240 | 0.10 | 0.19 | 0.00 | 1.00 | – | | | | |
| **Num Messages (log)** | 240 | 1.50 | 1.90 | 0.00 | 7.24 | 0.76*** | – | | | |
| **Score** | 44 | 3.89 | 0.56 | 2.63 | 4.88 | 0.38* | 0.53*** | – | | |
| **Num Submit** | 240 | 0.19 | 0.43 | 0.00 | 3.00 | 0.24*** | 0.30*** | -0.02 | – | |
| **Round** | 240 | 2.96 | 1.40 | 1.00 | 5.00 | -0.28*** | -0.30*** | -0.09 | 0.01 | – |

Note. M = mean; SD = standard deviation. Values in brackets are 95% confidence intervals.

The confidence interval is a plausible range of population correlations that could have produced the sample correlation (Cumming, 2014).

* p < .05, ** p < .01, *** p < .001.

**Table 1.** Means, standard deviations, and correlations of team-level variables with 95% CIs.

## Main results

### *Estimating attention responsiveness*

We then move from the statistical properties of the 306 distributed-lag models to examine the behavioral architecture they reveal. Each model estimates how the count of a given action type *A* at time *t* predicts the count of a response action *B* at time *t+1*, within team and grant period. Before turning to the latent factor, we first characterize the community-wide fixed-effect estimates to assess which action-response couplings are most reliably structured across the platform as a whole.

Figure 4 summarizes the statistically significant action-response relationships ($p < 10^{-8}$) as a directed network. The resulting structure reveals a coherent behavioral architecture in which targeted interpersonal signal (i.e. dyadic mentions, first direct interactions, and onboarding cues) propagate into downstream messaging and reactions at the platform level. More diffuse signals, such as channel-wide announcements or URL shares, generate comparatively weaker and less consistent downstream responses. This architecture is consistent with the theoretical view that attention direction operates through directed, person-to-person behavioral cues rather than broadcast communication and provides face validity for the action types that load most strongly on the ADA factor in subsequent analyses.

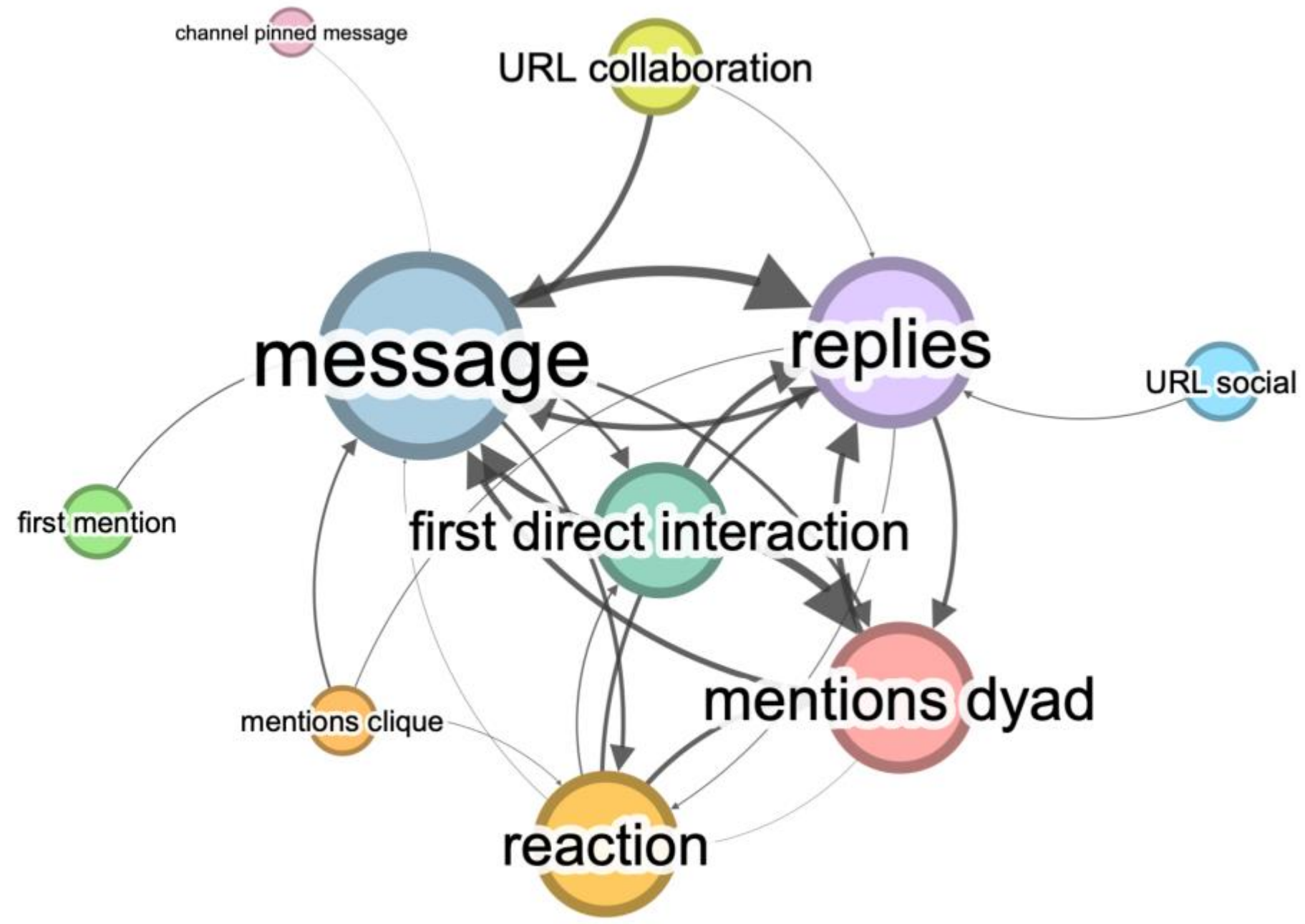

**Figure 4. Causal action–response network.** The network summarizes statistically significant action–response dynamics inferred from Slack interaction data. Nodes represent distinct action types, and directed edges indicate cases where activity in action $A$ at time $t$ significantly predicts increased activity in action $B$ at $t + 1$, based on distributed-lag negative-binomial mixed-effects models. The network displays fixed-effect estimates only and is restricted to edges with $p < 10^{-8}$ and to nodes with nonzero degree. Edge thickness indicates larger estimated causal effects, and node size is proportional to weighted degree.

## *A latent dimension of attention direction*

We now turn to the team-specific random slopes—the within-team, within-period deviations that capture how strongly each team's signals mobilize responses relative to the platform average. The responsiveness estimates at this level are inherently high-dimensional, reflecting one slope per ordered action pair per team-period. To assess whether these patterns reflect a coherent capability rather than a collection of unrelated pairwise effects, we summarize the random slopes at the action level. For each team-period and each trigger action $A$, we compute the mean of its positive responsiveness coefficients across all response actions $B$. This yields an action-level profile capturing how effective each class of signals is at eliciting downstream activity. This reduces the high-dimensional slope space to an interpretable action-level profile for each team-period. Consistent with the existence of a shared underlying structure, responsiveness coefficients across action types are systematically related: pairwise correlations among action-specific slopes are predominantly positive, with a mean inter-slope correlation of *0.18* (Supplementary Fig. S3).

A principal component analysis of this responsiveness matrix reveals a strong dominant latent dimension underlying attention responsiveness (Figure 5). The first component accounts for roughly one quarter of the variance in attention-triggering effectiveness and clearly exceeds what would be expected under a shuffled null model, whereas subsequent components cluster near the random baseline. This structure indicates that, despite the diversity of interaction types, teams differ along a common underlying dimension in how effectively their signals are taken up.

On this basis, we fit a one-factor exploratory factor model and extract a latent score for each team-period. We interpret this factor as Attention Direction Ability (ADA). Teams high in ADA are those in which behavioral signals reliably mobilize responses; teams low in ADA are those in which signals frequently dissipate without eliciting follow-on engagement. Consistent with its interpretation as a stable team-level capability rather than a period-specific artifact of communication intensity, ADA exhibits meaningful temporal coherence across grant rounds—Spearman rank correlations between consecutive rounds range from $\rho = 0.46$ *to* $\rho = 0.84$, with stability increasing as teams accumulate shared experience over the course of the program (see Robustness checks, Temporal stability).

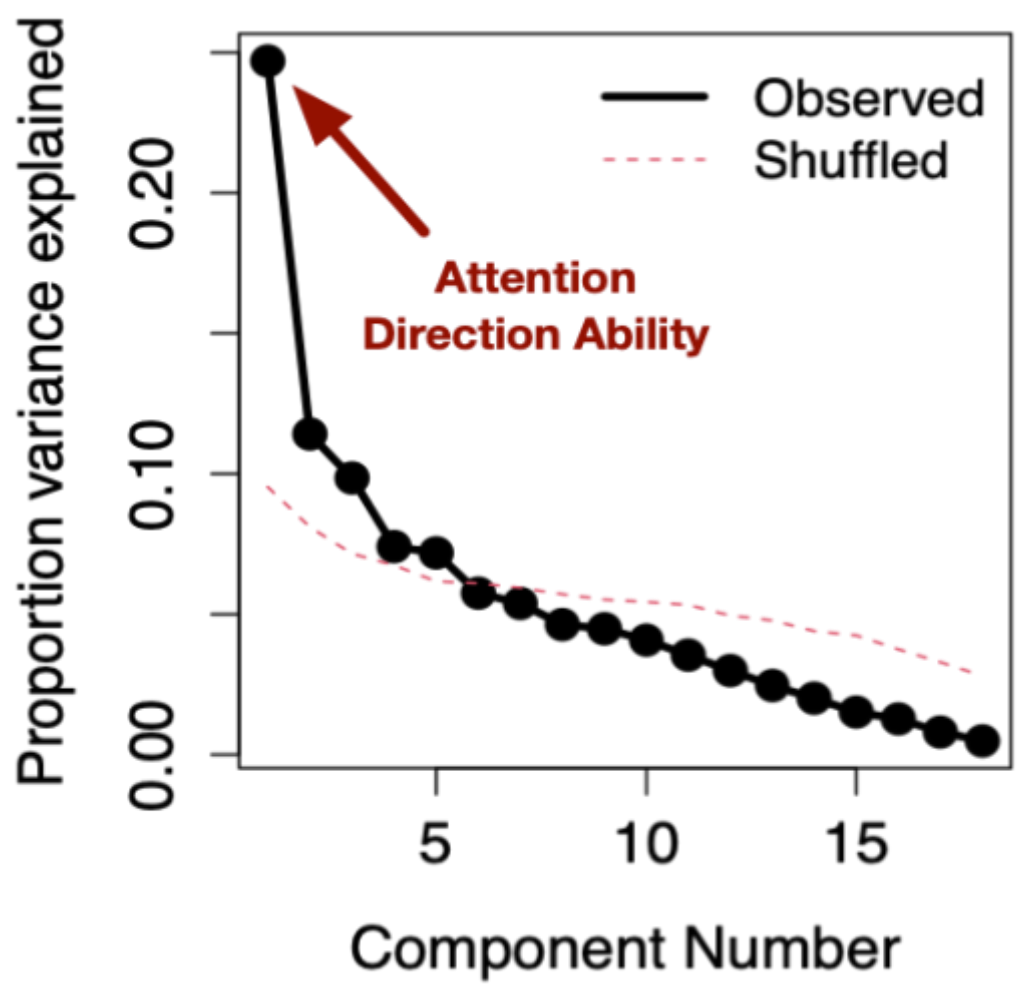

**Figure 5. Dominant latent dimension underlying attention responsiveness.** Principal component analysis (PCA) of the responsiveness matrix. The dashed curve shows the proportion of variance explained by successive principal components. The first component explains substantially more variance than subsequent components and exceeds a shuffled null benchmark (horizontal dashed line), indicating the presence of a dominant latent dimension.

### *Attention-Directing Ability predicts both submission and evaluation success*

If ADA reflects an underlying coordination capability, it should help explain both when teams participate in the grant process and how well they perform once they do. However, only a subset of project teams submits proposals in each grant round (Supplementary Figure S1), raising concerns about nonrandom sample selection. Unobserved factors that increase the likelihood of applying may also influence how evaluators score submitted proposals. To address this issue, we estimate a Heckman selection model (Tobit-2 specification) in which the selection equation models the binary decision to submit a proposal and the outcome equation models the evaluation score conditional on submission (Table 2 below).

Identification relies on an exclusion restriction based on project timing. Specifically, we use the number of days between a team's channel creation and the grant deadline as a predictor of submission that plausibly affects teams' opportunity and readiness to apply but does not directly influence evaluators' assessments. Consistent with this logic, a longer delay between channel creation and the deadline significantly reduces the likelihood of submission (*probit coefficient* $= -0.0020$, $p = .023$), while having no statistically significant association with evaluation scores among submitting teams (*p = .16*), satisfying standard criteria for a valid exclusion restriction.

| | Maximum Likelihood estimation | | 2-step Heckman / heckit estimation | |
|---|---|---|---|---|
| **Variables** | **Coef.** | **S.E.** | **Coef.** | **S.E.** |
| **Selection equation:** | | | | |
| Intercept | -0.893*** | (0.139) | -0.923*** | (0.143) |
| Attention-Directing Ability | 1.422** | (0.432) | 1.450** | (0.436) |
| Days since creation | -0.002* | (0.001) | -0.002. | (0.001) |
| **Outcome equation:** | | | | |
| Intercept | 3.092*** | (0.490) | 2.159 | (1.313) |
| Attention-Directing Ability | 1.219** | (0.452) | 2.068* | (1.004) |
| **Error terms:** | | | | |
| σ (sigma) | 0.626*** | (0.174) | 1.095 | (NA) |
| ρ (rho) | 0.665* | (0.323) | 0.996 | (NA) |
| Inverse Mills ratio | | | 1.091 | (0.832) |

Notes: Standard errors in parentheses. N = 240 (196 censored, 44 observed). σ denotes the outcome equation error standard deviation; ρ denotes the correlation between selection and outcome errors. ML log-likelihood = −138.77. *** p < 0.001, ** p < 0.01, * p < 0.05, . p < 0.1

**Table 2. Selection-corrected effects of ADA on grant submission and evaluation score.** Estimates from a Heckman selection model (Tobit-2 specification) jointly modeling grant submission and evaluation score conditional on submission. Days since channel creation serves as an exclusion restriction. Standard errors in parentheses.

Both full-information maximum likelihood and two-step estimators yield consistent results. The model detects meaningful nonrandom selection: the correlation between unobserved determinants of submission and scoring is positive and significant ($\rho = 0.66$, $p = .04$), indicating that teams who enter the competition are systematically advantaged in ways relevant for performance. This significant $\rho$ confirms the need to correct for selection bias—standard regressions would otherwise misattribute these latent advantages to observable predictors.

After accounting for selection, we find that the latent ADA strongly predicts both the probability of submitting ($\beta = 1.42$, $p = .001$) and teams' conditional performance ($\beta = 1.22$, $p = .0075$). Descriptive relationships are shown in *Supplementary Figure S4*. These results indicate that attention-direction ability shapes both entry into the competition and the success of teams who choose to participate.

## *Causal mediation analysis*

To clarify the process through which ADA translates into performance, we examine whether communication effort (measured as message volume) mediates the relationship between ADA and grant outcomes. ADA captures an emergent coordination capability: the extent to which signals within a channel reliably elicit subsequent engagement from others. A key implication of this view is that attention-direction ability should operate, at least in part, by mobilizing sustained collective effort, which in turn shapes participation and performance. To test this pathway, we estimate separate mediation models for submission (N = 240) and score (N = 44), controlling for grant rounds (Figure 6).

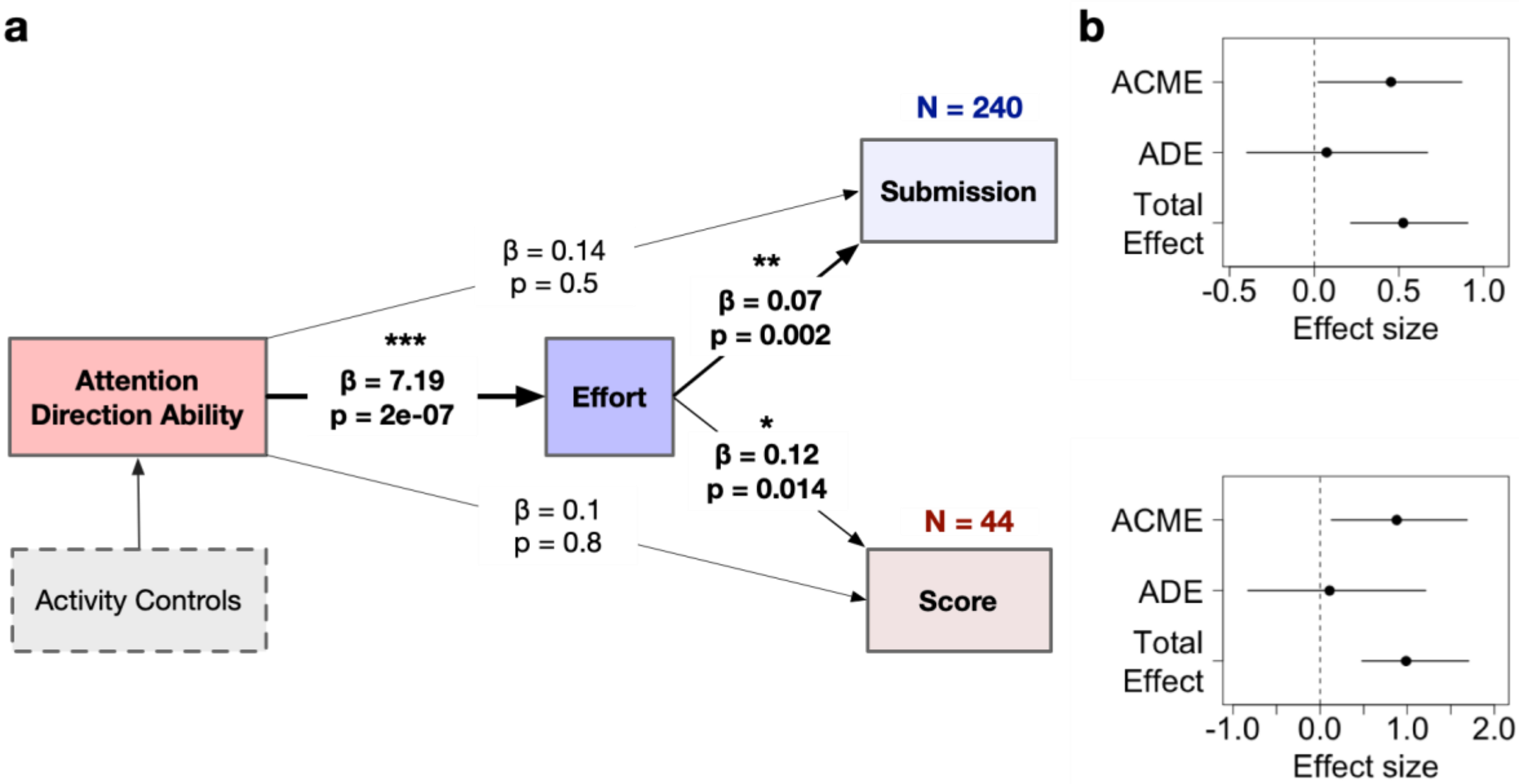

**Figure 6. Communication effort mediates the effect of ADA on submission and evaluation score.** Panel (a) shows estimated path coefficients linking ADA to communication effort (message volume) and downstream outcomes, controlling for grant round and activity controls. Panel (b) reports causal mediation estimates for submission (top; N = 240) and evaluation score conditional on submission (bottom; N = 44). Points denote point estimates and horizontal bars indicate 95% bootstrap confidence intervals (1,000 resamples) for the average causal mediation effect (ACME), average direct effect (ADE), and total effect. For submission, the indirect effect through communication effort is significant (*ACME = 0.48, 95% CI [0.06, 0.93], p = 0.026*), whereas the direct effect is not (*ADE = 0.14, 95% CI [–0.32, 0.69]*). For evaluation score, the mediated effect is likewise significant (*ACME = 0.88, 95% CI [0.13, 1.69], p = 0.026*), with a non-significant direct effect (*ADE = 0.11, 95% CI [–0.83, 1.21]*). A large share of ADA's total effect is mediated by communication effort (77% for submission; 89% for score).

For submission, ADA has a sizeable total effect on the probability that a team submits a proposal (*Total effect = 0.62, 95% CI [0.26, 1.03],* $p < 2 \times 10^{-16}$). Most of this influence is transmitted through message volume: the average causal mediation effect (ACME) is positive and significant (*ACME = 0.48, 95% CI [0.06, 0.93], p = 0.026*), whereas the residual direct effect of ADA after accounting for messaging is small and not significant (*ADE = 0.14, 95% CI [–0.32, 0.69], p = 0.56*). The estimated proportion mediated is 0.77, indicating that roughly three quarters of ADA's effect on submission operates through increased messaging.

We observe a similar pattern for score conditional on submission. ADA has a positive total effect on evaluation scores (*Total effect = 0.99, 95% CI [0.48, 1.70],* $p < 2\times10^{-16}$), and this effect is again largely mediated by message volume (*ACME = 0.88, 95% CI [0.13, 1.69], p = 0.026*). The direct effect of ADA independent of messaging is small and non-significant (*ADE = 0.11, 95% CI [–0.83, 1.21], p = 0.81*), with an estimated 89% of the total effect mediated. Full mediation estimates are reported in Supplementary Tables S1 (score) and S2 (submission).

These analyses suggest that ADA's impact on both participation and performance operates primarily through increased effort: teams with stronger attention-directing ability tend to generate more effort, and this heightened engagement, in turn, is what makes teams more likely to submit grant proposals and to receive higher scores. One concern is that ADA might simply be confounded by overall activity volume. A sensitivity analysis under the sequential ignorability framework (Imai, Keele, Tingley, et al., 2010) rules this out for score: the mediated effect remains positive unless an unmeasured confounder correlates at least $\rho = 0.4$ with both mediator and outcome residuals simultaneously—implying it would explain 16% of residual variance in each equation, far beyond any plausible omitted variable. The threshold is lower for submission ($\rho = 0.2$), consistent with the fact that crossing the participation bar is more directly tied to raw activity volume than achieving a high evaluation score; in both cases the direct effect is negligible at $\rho = 0$, supporting full mediation through effort. Sensitivity plots are reported in Supplementary Figure S5.

### ***Robustness checks***

***Model specification***. We explore the robustness of our findings to alternative modeling decisions and find substantial agreement across specifications (Supplementary Tables S4–S6; Figures S6–S7). Specifically, we examine (a) alternative data inclusion windows, (b) alternative temporal aggregation schemes, and (c) alternative latent factor structures for ADA. First, to account for the fact that coordination often begins before formal channel creation, we re-estimate all models while including activity in the one-month period preceding channel creation. This substantially increase the data used to estimate attention dynamics (*141,754 actions aggregated into 669,659 bins*). Under this specification, both selection into submission and conditional performance effects of ADA remain positive and statistically significant, and mediation patterns are preserved (Table S4; Figure S6, top row). Second, we assess sensitivity to temporal aggregation by computing attention dynamics using 1-day rather than 1-hour bins. Despite coarser temporal resolution, ADA continues to predict submission and evaluation outcomes with comparable magnitudes, and mediation effects remain directionally consistent (Table S5; Figure S6, bottom row). Third, we examine whether results depend on the dimensionality of the attention construct by replacing the single-factor ADA measure with a three-factor exploratory factor solution. Across specifications, the first factor consistently emerges as the dominant component, capturing the largest share of variance and exhibiting the strongest and most stable selection and mediation effects. While the remaining factors display weaker and more variable associations with outcomes, indirect (Average causal mediation effect (ACME)) and total effects remain positive across all factors (Table S6; Figure S7). Together, these analyses indicate that the core selection and mediation mechanisms are robust to alternative latent representations, even as different attention architectures exhibit distinct effect profiles.

***Construct operationalization***. Beyond model specification, we examined whether the positive-slope filter applied in Step 2, which isolates upward mobilization dynamics, is a theoretically necessary feature of the construct or an arbitrary modeling choice. Comparing the baseline specification to net responsiveness (sum of all slopes regardless of sign) and total interaction magnitude (sum of absolute slope values), factor loading vectors remain substantially correlated across specifications ($r = 0.83$ and $r = 0.81$ respectively), but predictive validity diverges: net responsiveness shows no significant association with either outcome, while total magnitude replicates the Heckman results but loses the mediation for score. The baseline specification is the only one that jointly identifies both the performance effect and the effort-mediation mechanism, consistent with the interpretation that what distinguishes high-ADA teams is how effectively their signals mobilize goal-directed engagement rather than how reactive they are in general. Full results are reported in Supplementary Section "Sensitivity to slope aggregation rule".

***Temporal stability***. Beyond sensitivity to modeling and operationalization choices, we examined whether ADA exhibits the temporal coherence expected of a latent team-level capability rather than a period-specific artifact of communication intensity. Specifically, because ADA is estimated separately for each team–grant-period observation, each team that participates across multiple rounds yields multiple independent ADA scores allowing us to assess whether these scores are consistent across periods, indicating whether a team's relative standing in attention-directing ability in one round predicts its standing in subsequent rounds. Spearman rank correlations between consecutive rounds range from $\rho = 0.46$ to $\rho = 0.84$, indicating that the relative ordering of teams by attention-directing ability is highly preserved and increasingly so as the program unfolds. *Lag-1* and *lag-2* Pearson correlations are $r = 0.32$ $(p < .001)$ and $r = 0.24$ $(p = .014)$ respectively, confirming that ADA in one round carries meaningful predictive information about ADA in subsequent rounds. The strengthening of stability over time is consistent with prior work showing that latent collective capabilities are more reliably expressed in established groups than in novel ones (Kommol et al., 2025). Full results, including a discussion of potential survivor bias in the increasing stability estimates, are reported in Supplementary Section "Temporal stability of Attention-Directing Ability".

## CONTRIBUTION AND DISCUSSION

In this research, we examine how teams mobilize and direct attention to achieve their collective goals in digitally mediated collaboration. This study explains why some distributed teams successfully mobilize collective action and achieve superior performance. While others struggle, even when they have similar access to the tools and infrastructure. We show that the difference is not only how much teams communicate, but how effectively communication cues trigger responses. In particular, we see that differences in team outcomes depend on a latent coordination capability embedded in interaction dynamics themselves. We conceptualize this capability as ADA: the extent to which everyday behavioral signals reliably trigger downstream engagement, sustain collective effort and shape trajectories of collaborative work.

Using high-resolution digital trace data from the OpenCovid19 initiative, we model the responsiveness structure among interaction signals and demonstrate that these interaction dynamics collapse onto a dominant latent factor that predicts both participation and expert-evaluated performance. Importantly, selection-corrected models show that ADA predicts outcomes beyond non-random entry into the grant process. Mediation analyses further show that ADA affects both submission and performance mainly through collective effort, measured as message volume. In other words, ADA operates as an upstream coordination capability: it shapes whether teams can sustain engagement long enough to produce evaluated outputs.

### *Attention-Directing Ability*

We advance research on collective attention by reconceptualizing attention not as a static distribution of cognitive focus, but as an emergent relational capability embedded in interaction dynamics. Existing work has primarily conceptualized collective attention either as the amount of attention allocated toward an issue or a task (Lorenz-Spreen et al., 2019; De Domenico and Altmann, 2020), or as co-attention and synchronized focus where individuals are attending to the same object at the same time (Shteynberg, 2015; Woolley et al., 2023). While these approaches have demonstrated that attention matters for collective intelligence and coordination, they leave underexplored how teams direct one another's attention in ongoing interaction.

Our findings indicate that teams with higher ADA perform better. Prior research has linked team performance to inputs such as expertise, diversity, and structure, as well as to shared cognitive states like collective intelligence and shared mental models (Woolley et al., 2010; Mathieu et al., 2019; Chikersal et al., 2017). We extend this literature by showing that what differentiates higher performing teams is not simply how much attention they allocate, but how effectively attention-directing signals translate into coordinated action. ADA captures this effectiveness directly, by measuring whether everyday interaction cues reliably elicit responses from others. Our findings indicate that teams with higher ADA are significantly more likely to submit grant proposals and, conditional on submission, to receive higher expert evaluation scores. These effects remain robust after correcting for non-random selection into participation, indicating that ADA captures whether behavioral signals, such as mentions, replies, onboarding cues, coordination messages, successfully recruit subsequent engagement from others. In doing so, we extend the attention-based view of organizing (Kauppila & Vaara, 2026; Ocasio, 1997; Ocasio et al., 2018), by showing that attention is not simply a scarce cognitive resource distributed across issues, but a dynamic interactional process through which teams recursively generate coordinated engagement. Attention direction thus becomes a property of collective organizing itself rather than merely an aggregate of individual focus, and this emergent direction constitutes a measurable coordination capability.

This perspective also complements recent work on collective intelligence (Woolley et al., 2010; Mayo and Woolley, 2021; Kommol et al., 2025) by identifying a concrete interactional mechanism through which collective capabilities emerge. Research on collective intelligence established that teams possess a general latent ability (Riedl et al., 2021) that predicts performance across diverse tasks, and that this ability is associated with equality of conversational turn-taking and members' social sensitivity (Woolley et al., 2010). However, these compositional correlates describe conditions under which collective intelligence arises, not the coordination process through which it operates. Attention-directing ability offers a process-level explanation. Social sensitivity is what allows members to direct attention effectively—inferring what a teammate is attending to, knows, and needs, so that a mention, question, or proposal is well-targeted rather than misfires—while equal turn-taking ensures such bids are distributed across members and can be picked up rather than monopolized. That is, they are enabling conditions for effective attention direction. What the collective intelligence factor captures as a statistical regularity, ADA specifies as a dynamic mechanism: the micro-level responsiveness structure through which teams convert individual signals into collective engagement. Importantly, ADA is a real latent ability: it is persistent and stable over time, as evidenced by Spearman rank correlations between consecutive rounds ranging from $\rho = 0.46$ to $\rho = 0.84$ and increasing as teams accumulate shared experience. This rules out the alternative that ADA merely captures a productive episode or a fortunate composition of members in a single period, and positions it as a genuine team-level coordination capability that characterizes how a team operates across the full arc of its collaborative work. This contributes to the literature on coordination by showing how it can be identified as a stable, emergent property of interaction dynamics that unfolds across the full trajectory of team collaboration (Faraj & Sproull, 2000; Okhuysen & Bechky, 2009).

### *Team emergence and process dynamics*

This study contributes to process-oriented and complexity perspectives on teamwork by showing how latent team capabilities emerge from temporal interaction dynamics rather than from static properties. Team research has increasingly emphasized emergence, temporality, and nonlinear interaction dynamics (Cronin et al., 2011; Humphrey and Aime, 2014; Mathieu et al., 2019). Yet empirical operationalizations often remain tied to aggregate measures such as communication frequency, network density, or shared cognition.

More specifically, we contribute to the literature by introducing a methodological and theoretical framework capable of linking micro-level interaction sequences to macro-level collective capabilities. ADA is not inferred from self-reports or aggregate communication counts. It emerges from the causal responsiveness structure among behavioral signals over time. This enables us to distinguish teams in which interaction cues reliably cascade into coordinated activity from those in which signals dissipate locally. In this sense, our study advances theories of team emergence by conceptualizing coordination capability as a latent relational architecture unfolding dynamically through interaction. This resonates with process views of organizing (Langley et al., 2013; Hernes, 2014), which emphasize that organizational outcomes emerge through ongoing flows of activity rather than stable structures alone. ADA captures this processual dimension: the recursive capacity of teams to transform local signals into sustained collective trajectories.

Methodologically, this study introduces an approach that links micro-level interaction sequences to macro-level team constructs without collapsing temporal and directional structure into aggregate statistics. By estimating causal responsiveness among behavioral signals and extracting a latent factor from the resulting profiles, the approach preserves who triggered what and distills it into a team-level measure with a clear causal interpretation. This bridges a persistent gap between the high-resolution digital trace data increasingly available in organizational settings and the team-level constructs that theory operates on.

### *Implications for future research*

This research opens several avenues for future research on collective attention, coordination, and digitally mediated organizing. First, future work could further investigate the temporal emergence and evolution of attention-direction capabilities. While our findings suggest that ADA exhibits meaningful stability across project phases, attention capabilities are unlikely to be entirely fixed. Teams may learn to direct attention more effectively over time through repeated interaction, leadership interventions, or the development of shared routines. Longitudinal process research could therefore examine how attention architectures evolve, stabilize, or decay across different stages of collaboration and under varying environmental conditions.

Second, our study invites further theorization of attention as a relational and distributed capability rather than an individual cognitive resource. Existing research has often conceptualized attention at the level of individuals, shared mental models, or issue salience (Ocasio 1997; Shteynberg 2015). By contrast, our findings suggest that collective attention emerges through interactional responsiveness structures that recursively shape engagement trajectories. Future research could investigate how attention-direction capabilities interact with other collective capabilities, such as collective memory, reasoning, psychological safety, or transactive knowledge systems (Woolley et al., 2023; Kommol et al., 2025). This would contribute to a more integrated understanding of how multiple emergent processes jointly shape collective intelligence and organizational adaptation.

Third, our findings create opportunities for advancing research on virtual work and digitally mediated collaboration. As organizations increasingly rely on remote, hybrid, and platform-based forms of work, coordination depends less on co-presence and more on digitally mediated interaction dynamics (Raghuram et al., 2019; Yang et al., 2022). Future research could examine how technological affordances shape teams' ability to direct attention under different collaboration architectures. For example, some platforms may amplify attentional fragmentation through notification overload, while others may better support the propagation of meaningful signals. Understanding how digital infrastructures enable or constrain attention direction could substantially deepen organizational research on collaboration technologies and distributed work.

Fourth, our study raises broader methodological opportunities for organizational scholarship. Organizational research has long emphasized the importance of emergence and process dynamics (Langley et al., 2013; Humphrey & Aime, 2014), yet empirical approaches often struggle to connect micro-level interactions with higher-order organizational constructs. By leveraging high-resolution behavioral traces to estimate latent coordination capabilities, our approach demonstrates one pathway for operationalizing emergent organizational processes. Future research could extend similar approaches to study leadership emergence, resilience, creativity, conflict dynamics, or institutional coordination in digitally mediated environments.

Finally, future research could investigate the boundary conditions of attention-direction capability. ADA may operate differently depending on task complexity, temporal pressure, team composition, leadership structure, or the degree of interdependence among members. In highly routinized settings, strong attention-direction may reinforce efficiency and execution, whereas in exploratory contexts it may facilitate experimentation and adaptive search. Understanding when and how attention-direction capabilities contribute to coordination, innovation, or resilience would significantly extend both the attention-based view of organizing and contemporary theories of collective capability.

### *Implications for practice*

Our work also has practical implications. Traditional coordination relies on roles, routines, and reporting lines, but underneath these structures lie the attentional dynamics that determine what gets noticed and acted upon. Managers and platform designers can enhance team effectiveness by deliberately shaping attention direction: for example, by introducing clear prioritization cues early in projects, reinforcing focal points as work unfolds, and designing digital environments that highlight consequential signals while reducing noise (Bernstein et al., 2024; Faraj et al., 2011). Instead of focusing solely on increasing communication volume, organizations could explore how signals propagate and mobilize collective responses.

More broadly, our study suggests that organizations should treat attentional coordination as a strategic organizational capability rather than as an incidental byproduct of communication. In remote and hybrid environments, teams frequently operate with fluid membership, asynchronous interaction, and reduced opportunities for informal coordination. Under such conditions, organizations that more effectively direct collective attention may gain substantial advantages in adaptability, innovation, and execution.

Finally, our work suggests that organizations may benefit from developing metrics and diagnostics capable of identifying emergent coordination patterns in real time. Rather than relying solely on lagging indicators of performance, organizations could monitor how interaction signals propagate, whether engagement is sustained, and where coordination bottlenecks emerge. Such approaches may help organizations intervene earlier in struggling collaborations and better support collective problem solving in increasingly complex and distributed forms of work.

## REFERENCES

Ancona, D. G., & Chong, C. L. (1999). *Cycles and synchrony: The temporal role of context in team behavior*.

Arrow, H., McGrath, J. E., & Berdahl, J. L. (2000). *Small groups as complex systems: Formation, coordination, development, and adaptation*. Sage Publications.

Bailey, D. E., Faraj, S., Hinds, P. J., Leonardi, P. M., & von Krogh, G. (2022). We Are All Theorists of Technology Now: A Relational Perspective on Emerging Technology and Organizing. *Organization Science*, *33*(1), 1–18. https://doi.org/10.1287/orsc.2021.1562

Bates, D., Mächler, M., Bolker, B., & Walker, S. (2015). Fitting linear mixed-effects models using lme4. *Journal of Statistical Software*, *67*, 1–48.

Bernstein, E. S., Gupta, P., Mortensen, M., & Leonardi, P. M. (2024). Collective attention and relational overload: A theory of transactive control in high-permeability intraorganizational environments. *Research in Organizational Behavior*, *44*, 100209.

Blay, T., Froese, F. J., Taras, V., & Gunkel, M. (2024). Convergence of collaborative behavior in virtual teams: The role of external crises and implications for performance. *Journal of Applied Psychology*, *109*(4), 469.

Bloom, N., Han, R., & Liang, J. (2024). Hybrid working from home improves retention without damaging performance. *Nature*, *630*(8018), 920–925.

Butterworth, G., & Jarrett, N. (1991). What minds have in common is space: Spatial mechanisms serving joint visual attention in infancy. *British Journal of Developmental Psychology*, *9*(1), 55–72.

Chikersal, P., Tomprou, M., Kim, Y. J., Woolley, A. W., & Dabbish, L. (2017). Deep structures of collaboration: Physiological correlates of collective intelligence and group satisfaction. *Proceedings of the 2017 ACM Conference on Computer Supported Cooperative Work and Social Computing*, 873–888.

Cramton, C. D. (2001). The mutual knowledge problem and its consequences for dispersed collaboration. *Organization Science*, *12*(3), 346–371.

Cronin, M. A., Weingart, L. R., & Todorova, G. (2011). Dynamics in groups: Are we there yet? *Academy of Management Annals*, *5*(1), 571–612.

De Domenico, M., & Altmann, E. G. (2020). Unraveling the origin of social bursts in collective attention. *Scientific Reports*, *10*(1), 4629.

DeChurch, L. A., & Mesmer-Magnus, J. R. (2010). The cognitive underpinnings of effective teamwork: A meta-analysis. *Journal of Applied Psychology*, *95*(1), 32.

Faraj, S., Jarvenpaa, S. L., & Majchrzak, A. (2011). Knowledge collaboration in online communities. *Organization Science*, *22*(5), 1224–1239.

Faraj, S., & Sproull, L. (2000). Coordinating expertise in software development teams. *Management Science*, *46*(12), 1554–1568.

Feldman, M. S., & Pentland, B. T. (2003). Reconceptualizing organizational routines as a source of flexibility and change. *Administrative Science Quarterly*, *48*(1), 94–118.

Felin, T., Foss, N. J., Heimeriks, K. H., & Madsen, T. L. (2012). Microfoundations of routines and capabilities: Individuals, processes, and structure. *Journal of Management Studies*, *49*(8), 1351–1374.
Furst, S. A., Reeves, M., Rosen, B., & Blackburn, R. S. (2004). Managing the life cycle of virtual teams. *Academy of Management Perspectives*, *18*(2), 6–20.
Gelman, A., & Hill, J. (2007). *Data analysis using regression and multilevel/hierarchical models*. Cambridge university press.
Gibson, C. B., Dunlop, P. D., Majchrzak, A., & Chia, T. (2022). Sustaining effectiveness in global teams: The coevolution of knowledge management activities and technology affordances. *Organization Science*, *33*(3), 1018–1048.
Graham, C. L., Landrain, T. E., Vjestica, A., Masselot, C., Lawton, E., Blondel, L., Haenal, L., Tzovaras, B. G., & Santolini, M. (2023). Community review: A robust and scalable selection system for resource allocation within open science and innovation communities. *F1000Research*, *11*, 1440.
Greene, W. H. 2012. Econometric Analysis, 7th edition. Prentice-Hall, Upper Saddle River, NJ.
Hackman, J. R. (2012). From causes to conditions in group research. *Journal of Organizational Behavior*, *33*(3), 428–444.
Hackman, J. R., & Morris, C. G. (1975). Group tasks, group interaction process, and group performance effectiveness: A review and proposed integration. *Advances in Experimental Social Psychology*, *8*, 45–99.
Heckman, J. J. (1979). Sample selection bias as a specification error. *Econometrica: Journal of the Econometric Society*, 153–161.
Helfat, C. E., & Peteraf, M. A. (2003). The dynamic resource-based view: Capability lifecycles. *Strategic Management Journal*, *24*(10), 997–1010.
Hernes, T. (2014). *A process theory of organization*. Oup Oxford.
Humphrey, S. E., & Aime, F. (2014). Team microdynamics: Toward an organizing approach to teamwork. *Academy of Management Annals*, *8*(1), 443–503.
Ilgen, D. R., Hollenbeck, J. R., Johnson, M., & Jundt, D. (2005). Teams in organizations: From input-process-output models to IMOI models. *Annu. Rev. Psychol.*, *56*(1), 517–543.
Imai, K., Keele, L., & Tingley, D. (2010). A general approach to causal mediation analysis. *Psychological Methods*, *15*(4), 309.
Imai, K., Keele, L., Tingley, D., & Yamamoto, T. (2010). Causal mediation analysis using R. In *Advances in social science research using R* (pp. 129–154). Springer.
Janicik, G. A., & Bartel, C. A. (2003). Talking about time: Effects of temporal planning and time awareness norms on group coordination and performance. *Group Dynamics: Theory, Research, and Practice*, *7*(2), 122.
Jeppesen, L. B., & Lakhani, K. R. (2010). Marginality and problem-solving effectiveness in broadcast search. *Organization Science*, *21*(5), 1016–1033.
Kauppila, O.-P., & Vaara, E. (2026). Triadic Attentional Processing and Buildup of Joint Attentional Engagement in Emergent Strategic Initiatives. *Academy of Management Review*, (ja), amr. 2024.0120.
Klein, K. J., & Kozlowski, S. W. (2000). From micro to meso: Critical steps in conceptualizing and conducting multilevel research. *Organizational Research Methods*, *3*(3), 211–236.
Kokshagina, O. (2021). Open Covid-19: Organizing an extreme crowdsourcing campaign to tackle grand challenges. *R&D Management*.

Kommol, E., Riedl, C., & Woolley, A. (2025). The structure of collective intelligence: Evidence for collective memory, attention, and reasoning. OSF Preprints. https://osf.io/4sqfx/

Kossek, E. E., Gettings, P., & Misra, K. (2021). The future of flexibility at work. *Harvard Business Review*, *28*.

Kossek, E. E., Perrigino, M. B., & Lautsch, B. A. (2023). Work-life flexibility policies from a boundary control and implementation perspective: A review and research framework. *Journal of Management*, *49*(6), 2062–2108.

Kozlowski, S. W. (2025). A multilevel, emergent journey to unpack team process dynamics. *Small Group Research*, *56*(3), 487–523.

Langley, A. N. N., Smallman, C., Tsoukas, H., & Van de Ven, A. H. (2013). Process studies of change in organization and management: Unveiling temporality, activity, and flow. *Academy of Management Journal*, *56*(1), 1–13.

Lix, K., Goldberg, A., Srivastava, S. B., & Valentine, M. A. (2022). Aligning differences: Discursive diversity and team performance. *Management Science*, *68*(11), 8430–8448.

Lorenz-Spreen, P., Mønsted, B. M., Hövel, P., & Lehmann, S. (2019). Accelerating dynamics of collective attention. *Nature Communications*, *10*(1), 1759.

Mathieu, J. E., Gallagher, P. T., Domingo, M. A., & Klock, E. A. (2019). Embracing complexity: Reviewing the past decade of team effectiveness research. *Annual Review of Organizational Psychology and Organizational Behavior*, *6*(1), 17–46.

Montoya-Weiss, M. M., Massey, A. P., & Song, M. (2001). Getting it together: Temporal coordination and conflict management in global virtual teams. *Academy of Management Journal*, *44*(6), 1251–1262.

Ocasio, W. (1997). Towards an attention-based view of the firm. *Strategic Management Journal*, *18*(S1), 187–206.

Ocasio, W., Laamanen, T., & Vaara, E. (2018). Communication and attention dynamics: An attention-based view of strategic change. *Strategic Management Journal*, *39*(1), 155–167.

Okhuysen, G. A., & Bechky, B. A. (2009). 10 coordination in organizations: An integrative perspective. *Academy of Management Annals*, *3*(1), 463–502.

Raghuram, S., Hill, N. S., Gibbs, J. L., & Maruping, L. M. (2019). Virtual work: Bridging research clusters. *Academy of Management Annals*, *13*(1), 308–341.

Riedl, C., De Cremer, D., Lucarelli, G., Antoine-Souklaye, E., Bullock, S., Ajmeri, N., Batty, M., Black, M., Cartlidge, J., & Challen, R. (2025). The potential and challenges of AI for collective intelligence. *Collective Intelligence*, *4*(1), 26339137241308821.

Riedl, C., Kim, Y. J., Gupta, P., Malone, T. W., & Woolley, A. W. (2021). Quantifying collective intelligence in human groups. *Proceedings of the National Academy of Sciences*, *118*(21), e2005737118. https://doi.org/10.1073/pnas.2005737118

Riedl, C., & Woolley, A. W. (2017). Teams vs. Crowds: A Field Test of the Relative Contribution of Incentives, Member Ability, and Emergent Collaboration to Crowd-Based Problem Solving Performance. *Academy of Management Discoveries*, *3*(4), 382–403. https://doi.org/10.5465/amd.2015.0097

Rosas, F. E., Mediano, P. A., Jensen, H. J., Seth, A. K., Barrett, A. B., Carhart-Harris, R. L., & Bor, D. (2020). Reconciling emergences: An information-theoretic approach to identify causal emergence in multivariate data. *PLoS Computational Biology*, *16*(12), e1008289.

Santolini, M. (2020). Covid-19: The rise of a global collective intelligence? The Conversation. http://theconversation.com/covid-19-the-rise-of-a-global-collective-intelligence-135738

Shteynberg, G. (2015). Shared attention. *Perspectives on Psychological Science*, *10*(5), 579–590.
Skiadopoulou, A., Gupta, V. K., Lee, Y., & Srivastava, S. (2026). "May we have your attention, please?": A bibliometric analysis of attention-based view research. *Journal of Management History*, 1–29.
Teece, D. J. (2007). Explicating dynamic capabilities: The nature and microfoundations of (sustainable) enterprise performance. *Strategic Management Journal*, *28*(13), 1319–1350.
Teece, D. J., Pisano, G., & Shuen, A. (1997). Dynamic capabilities and strategic management. *Strategic Management Journal*, *18*(7), 509–533.
Tomprou, M., Kim, Y. J., Chikersal, P., Woolley, A. W., & Dabbish, L. A. (2021). Speaking out of turn: How video conferencing reduces vocal synchrony and collective intelligence. *PLoS One*, *16*(3), e0247655.
Toomet, O., & Henningsen, A. (2008). Sample selection models in R: Package sampleSelection. *Journal of Statistical Software*, *27*, 1–23.
Trevarthen, C., & Aitken, K. J. (2001). Infant intersubjectivity: Research, theory, and clinical applications. *The Journal of Child Psychology and Psychiatry and Allied Disciplines*, *42*(1), 3–48.
Wooldridge, J. M. (2010). *Econometric analysis of cross section and panel data*. MIT press.
Woolley, A. W., Chow, R. M., Mayo, A. T., Riedl, C., & Chang, J. W. (2023). Collective attention and collective intelligence: The role of hierarchy and team gender composition. *Organization Science*, *34*(3), 1315–1331.
Yang, L., Holtz, D., Jaffe, S., Suri, S., Sinha, S., Weston, J., Joyce, C., Shah, N., Sherman, K., & Hecht, B. (2022). The effects of remote work on collaboration among information workers. *Nature Human Behaviour*, *6*(1), 43–54.

## SUPPLEMENTARY INFORMATION

### *Data extraction*

We analyzed digital trace data from the Slack workspace of the OpenCovid19 Initiative, a distributed open-innovation community launched in early 2020 in response to the COVID-19 pandemic. Using the workspace's event log, we obtained a time-stamped record of all relevant user actions across public channels. The raw dataset contains 36,951 unique events, identified by distinct timestamps, which expand to 86,046 coded actions once we decompose each event into its constituent behaviors (i.e., a message that both mentions a user and includes a URL generates multiple action codes). These actions were performed by 2,076 distinct sending users (sources) and involved a total of 2,233 unique users when including both senders and receivers (targets).

### *Mapping to Project Channels and Performance Outcomes*

From the entire Slack workspace, we isolated the subset of channels dedicated to project work. We identified 79 project-related channels, defined as channels created to coordinate specific OpenCovid19 projects and their associated discussions. Among these, 30 channels could be reliably mapped to downstream performance information from the Initiative's microgrant program, which records whether a team submitted a proposal and the reviewer score it received. Some teams submitted proposals in multiple grant rounds, producing 44 scored submissions across these 30 identifiable project channels. Restricting the trace to project channels yields 13,912 unique events and 29,218 coded actions, forming the core dataset used to study team-level attention dynamics.

### *Definition of Action Types*

To create a unified behavioral representation of project activity, each raw Slack event was re-coded into a discrete action type reflecting its functional contribution to coordination, attention allocation, or social interaction. Channel structuring behaviors were captured through the variables *channel_created, channel_pinned_message, channel_purpose_created, and channel_topic_created*. The *channel_created* action identifies the moment at which a project workspace becomes available to participants, while *channel_pinned_message* records the pinning of a message, a behavior that elevates specific content as collectively salient. The *channel_purpose_created* and *channel_topic_created* actions mark updates to the channel's purpose statement and topic header, respectively, which signal attempts to reframe or clarify the goals, scope, or current priorities of the project space.

Membership dynamics were recorded through *channel_joins*, which logs each instance of a user entering a project channel. This action captures the expansion of the active participant set and provides a temporal basis for distinguishing early engagement from late-stage participation.

Direct content contributions were represented by the message action, corresponding to the posting of a message in a project channel. Because a single message can embed several behavioral signals (including user mentions and URL references) it was decomposed into its constituent elements. Three complementary mention actions, *mentions_dyad, mentions_clique*, and *mentions_channel*, encode directed attention within the team. *Mentions_dyad* refers to the explicit tagging of a single user, *mentions_clique* captures messages that tag multiple specific users, and *mentions_channel* corresponds to channel-wide mentions such as @channel or @here, which broadcast urgency or request broad attention.

Responsiveness and feedback were represented through reaction and replies. The reaction action captures emoji reactions added to messages, which provide lightweight, low-friction forms of endorsement, acknowledgement, or emotional expression. The replies action identifies threaded responses, marking more substantive conversational engagement around a focal message.

Several actions correspond to community-wide events synchronized across the workspace. *Global_community_call* identifies announcements or reminders related to all-hands community meetings. *Microgrant_call* flags posts announcing the opening of a grant round, and *microgrant_result* marks the communication of funding decisions. These event actions reflect bursts of shared attention that may shape project focus or mobilization.

A further set of actions captures external reference behaviors through URL categories. The taxonomy distinguishes whether a link directs attention toward coding resources (*url_coding*), collaboration platforms (*url_collaboration*), event pages (*url_event*), JOGL platform resources (*url_jogl*), meeting links (*url_meeting*), news articles (*url_news*), scientific material (*url_science*), social media (*url_social*), or surveys (*url_survey*). Each URL category thus identifies a distinct class of resource that participants bring into the shared workspace, allowing the analysis to differentiate informational, scientific, organizational, and outreach-oriented forms of attention.

Finally, onboarding actions capture when a participant first becomes behaviorally visible in the dataset. *First_activity* records the user's earliest action of any type within the workspace, *first_direct_interaction* identifies the first instance in which the user interacts with another participant through a dyadic or clique mention, and first_mention marks the first time the user is referenced by someone else. These onboarding indicators allow us to align users temporally and model early engagement trajectories.

These action types provide a comprehensive representation of collaborative behavior in project channels. Each raw Slack event may produce multiple action codes, ensuring that all informational, structural, and social signals embedded in messages and interactions are systematically captured for subsequent modeling of attention dynamics and team performance.

### *Estimation subset of action type*

While we code a broad taxonomy of actions to represent the full behavioral repertoire in Slack, not all action types occur frequently enough within project channels and time bins to support stable estimation of distributed-lag models for every ordered pair ($A \rightarrow B$). Accordingly, the main analyses restrict attention-triggering models to the subset of action types with sufficient within-channel variation and coverage across channel-periods to identify random-slope effects: *channel_joins, channel_pinned_message, message, mentions_channel, mentions_clique, mentions_dyad, reaction, replies, url_coding, url_collaboration, url_jogl, url_meeting, url_news, url_science, url_social, first_activity, first_direct_interaction, and first_mention*. Action types that were too sparse (e.g., channel purpose/topic updates, some URL categories, and other low-frequency events) were retained in the raw coding but excluded from the baseline estimation sample. In robustness checks, we broaden this set to include all coded action types and verify that the substantive conclusions are not sensitive to the baseline restriction (see Tables S4–S6 and Figures S6–S7).

### *Inter-Event Time Analysis*

To select the temporal resolution for modeling attention dynamics, we computed the empirical distribution of inter-event times (IETs) between successive Slack actions within each project channel (Figure S2). For every channel, events were ordered by timestamp, and the elapsed time between adjacent events was calculated in hours. All IETs across the 79 project channels were pooled into a single distribution. We binned the resulting values on a logarithmic scale and plotted their density on log–log axes to characterize the temporal spacing of activity.

The distribution was heavy-tailed, with short intervals dominating. The median inter-event time was on the order of minutes, while the mean was approximately 1.5 hours, reflecting a mix of dense bursts of activity interspersed with longer pauses. Based on this empirical rhythm, we adopted a 1-hour temporal bin for constructing the time-series–cross-section panel. This bin size lies just below the mean IET, ensuring that most causally connected interactions fall within the same window while still generating a sufficiently large number of non-empty bins to support the estimation of distributed-lag mixed-effects models. All downstream analyses, including action counts, lag construction, and random-slope extraction, used this 1-hour resolution.

### *Data Preparation and Construction of the Panel Dataset*

For each project channel, we first identified the sequence of microgrant deadlines relevant to that channel and used these timestamps to divide the channel's activity into discrete Grant periods, defined as the intervals between consecutive deadlines. Channel creation marked the start of the first period. To ensure that the temporal segmentation aligned with the evaluative horizon of each project, we truncated periods differently depending on whether the team submitted at least once. For channels that submitted at least one proposal, we retained only the periods from channel creation up to and including the last deadline at which the team submitted; all periods following that deadline were excluded. For channels that never submitted, we retained periods from channel creation up to the last deadline for which the channel displayed any activity, excluding subsequent empty periods. After determining the final period boundary for each channel, we assigned every event to the unique Grant period whose interval contained its timestamp. This procedure yielded clean, non-overlapping, period-specific event streams for every project channel included in the analysis.

To create a uniform temporal structure for distributed-lag modeling, we generated a contiguous sequence of one-hour bins spanning each Grant period. The period boundaries defined the start and end of each bin sequence, and every bin was assigned the corresponding period label. Applying this procedure across all channels produced a total of 508,656 one-hour bins, representing the full time-series–cross-section structure used in the estimation stage.

Each raw event was then assigned to a bin by rounding its timestamp down to the nearest hour. Within each Grant period, we counted the number of occurrences of every action type in every time bin, resulting in a long-format dataset in which each row corresponded to a Channel × Period × TimeBin × Action observation.

Because many bins contained no activity for many action types, we expanded the dataset into the full Cartesian product of Channel × Grant period × TimeBin × Action. For each cell in this grid, we recorded the observed count if any events occurred during that bin, or zero otherwise. Converting this structure into a wide format produced, for each bin, a synchronized vector of action counts. This complete grid ensured that distributed-lag models used every potential observation window, rather than only bins in which activity occurred, which is essential for identifying action-to-action triggering effects.

### *Global Platform Controls*

To account for platform-wide fluctuations in activity, we computed two global controls aligned to the same one-hour bins. First, we counted the number of distinct users who were active anywhere on the platform during each bin. Second, we counted all actions occurring platform-wide in that bin and recorded their total. These global variables were merged into the channel-level panel by matching on the common TimeBin index; doing so ensures that subsequent models can separate the effect of local attention dynamics from broader exogenous shifts in overall platform engagement.

Finally, we added two time-structured controls: day of the week and calendar month. Both variables were converted into ordered factors to allow flexible adjustment for weekly and seasonal cycles in activity. Each row of the resulting dataset represents a single channel-period-bin, containing a complete vector of action counts, the number of platform-wide active users, the total platform activity, the day and month indicators, and a unique identifier for each *Channel* × *Grant* period. This panel serves as the basis for constructing lagged variables and fitting the negative-binomial mixed-effects models used to estimate causal attention-triggering effects.

### *Lag Construction for Attention-Triggering Models*

To estimate whether a given attention-directing action at time t increases the likelihood of a downstream response at time $t + 1$, we constructed lagged datasets for every ordered pair of action types ($A \rightarrow B$). Starting from the complete one-hour panel, we grouped observations by Channel and Grant period, sorted them temporally within each group, and created a lagged predictor $LagA$, defined as the count of action $A$ in the preceding time bin. The contemporaneous count of action B served as the dependent variable $Y_B$. We retained only observations for which the lagged bin was present, thereby ensuring that the dataset contained uninterrupted one-hour windows suitable for distributed-lag analysis.

For each ($A \rightarrow B$) pair, the resulting dataset consisted of the lagged predictor, the outcome variable, and all exogenous controls aligned to the same one-hour structure: the platform-wide active users, the total platform activity, the day-of-week factor, and the month-of-year factor. All continuous predictors were log-transformed using the *log(x + 1)* transformation and standardized to unit variance. This ensured comparability across action types and prevented numerical instability in the mixed-effects estimation. The construction of these lagged datasets enables us to test, for each action pair, whether increases in action *A* reliably trigger subsequent increases in action *B,* while adjusting for exogenous temporal fluctuations and global platform activity.

### *Estimation of Distributed-Lag Mixed Models*

For each action pair ($A \rightarrow B$), we estimated a hierarchical negative-binomial distributed-lag model of the form:

$$Y_{B,i,t} \sim NB(\mu_{N,i,t}, \phi)$$

$$\begin{aligned} log\ (\mu_{B,i,t}) \ = \ & \beta_0 + \beta_1\, LagA_{t-1} + \gamma_1\, log\ (AU_t + 1) + \gamma_2\, log\ (TA_t + 1) + DoW_t \\ & + Month_t \\ & + u_{0,channel} + u_{0,channel-period} + u_{1,channel-period} LagA_{i,t-1} \end{aligned}$$

Here, $Y_{B,i,t}$ is the count of action $B$ in the current time bin, $LagA_{t-1}$ is the standardized count of action $A$ in the preceding bin, and $(AU_t, TA_t)$ represent platform-wide active users and total platform activity. The model includes random intercepts for channels and channel-periods to account for unobserved heterogeneity in communication intensity, and a random slope on $LagA_{i,t-1}$ that varies by $Channel \times Grant$ period. This random slope $u_{A,channel-period}$ quantifies the degree to which a particular team, during a particular Grant period, is more or less responsive to action $A$ as a trigger of downstream activity.

Models were estimated using maximum likelihood via *glmmTMB* with the *nbinom1* variance structure, which accommodates overdispersed count data while maintaining computational stability across hundreds of millions of potential parameter combinations. To ensure model quality, we applied several filters before estimation: action pairs were only fitted if both actions exhibited sufficient activity (≥20 total occurrences each), if action $A$ appeared in at least five non-zero bins, and if at least ten channel-period groups exhibited any activity for the pair. This filtering strategy avoids degenerate models while preserving all substantively meaningful pairs.

In cases where the random-slope model failed to converge, we estimated a fallback model containing only random intercepts. If this fallback model converged, we extracted the fixed effect of $LagA_t$ but assigned missing values to the random slopes; if it too failed, the pair was marked as unestimable. For all converged models, we extracted two components: (1) the fixed effect of $LagA_t$, indicating the global tendency of action $A$ to trigger action $B$ across the whole community, and (2) the random slope $u_{A,channel-period}$, representing the team-period–specific causal responsiveness.

These random slopes—one per action pair, for every *Channel* × *Grant* period—serve as the building blocks for the latent attention-direction factor estimated in the subsequent stage of analysis.

### *Construction of Action-Level Slope Indicators and Latent Factor Extraction*

To summarize teams' attention-triggering patterns at the action level, we first aggregated the random slopes from the distributed-lag models. For each *Channel* × *Grant-period* unit ("ChannelPeriod") and each trigger action type $A$, we computed the mean positive random slope of $A$ on all response actions $B$ estimated for that *ChannelPeriod*. Concretely, for every $A$ we averaged the positive random slopes from all fitted $(A \rightarrow B)$ models within that *ChannelPeriod*, yielding a single indicator that captures how strongly action $A$ tends to elicit downstream activity of any type in that team-period. This produced a *ChannelPeriod* × *Action matrix* in which rows index *ChannelPeriods* and columns index trigger actions $A$, with entries equal to the aggregated "$A$ triggers *" slope values. Actions with extremely sparse coverage were removed, and remaining columns were standardized (z-scored) to place indicators on a common scale.

To determine the dimensionality of the latent structure across these action-level slopes, we applied principal component analysis (PCA) to the standardized matrix. PCA was used solely as a diagnostic for the number of latent dimensions, not for factor-score estimation. We computed the proportion of variance explained by each principal component and compared the resulting eigenvalue spectrum to a null model obtained by independently permuting the entries within each action column (preserving marginal distributions while destroying cross-action correlations). In the observed data, the first principal component explained 25.1% of total variance, whereas the second and third components explained 9.0% and 8.2%, respectively, with all subsequent components below 6.5%. In the shuffled data, all components lay in a narrow band between 2.3% and 5.9%. Only the first observed component clearly exceeded this null envelope, while the remaining components fell close to the null profile. Parallel analysis based on this same matrix yielded the same conclusion. We therefore retained a single latent dimension.

Given this PCA-based dimensionality assessment, we then estimated a one-factor exploratory factor analysis (EFA) model on the same standardized action-level slope matrix. EFA, rather than PCA, was used for factor-score extraction because it explicitly partitions common variance from unique and residual variance. Using maximum-likelihood extraction with oblique rotation, we obtained regression-based factor scores for each ChannelPeriod. These scores constitute our measure of Attention-Direction Ability (ADA): a latent team-period–level capability that captures how effectively a team's various attention-directing actions, taken together, trigger subsequent activity in the group.

### *Grant-Period Regression of Performance on Attention-Direction Ability*

To evaluate whether teams with higher attention-direction capability perform better during the microgrant process, we estimated regression models linking the latent ADA factor to performance outcomes measured at the team-round level (*Channel* × *Grant-period*). In the OpenCovid19 program, teams may submit multiple proposals within the same grant round (e.g., revised versions or parallel submissions). Accordingly, each submission receives an independent reviewer score, and all scores for all submissions made by a team in a given round were retained in the dataset and included in the analysis.

For each team i and grant round r, we modeled two outcomes:

1. Submission count ($Submitted_{ir}$): the number of proposals the team submitted in round $r$;
2. Score ($Score_{isr}$): the reviewer score (0–5) assigned to each individual submission $s$ made by team $i$ in round $r$. If a team submitted $k$ proposals in a round, all $k$ score observations were included.

We first regressed the submission count on ADA using:

$$Submitted_{ir} = \alpha_0 + \alpha_1 ADA_{ir} + \delta_r + \epsilon_{ir}$$

where $\delta_r$ denotes round fixed effects that adjust for systematic differences across grant waves (e.g., changes in review load, call clarity, or project maturity).

We then estimated the association between ADA and submission-level scores through the model:

$$Score_{isr} = \beta_0 + \beta_1 ADA_{ir} + \delta_r + \eta_{ir}$$

where each submission inherits the ADA value of its corresponding team-round. This formulation ensures that all available evaluation data are used while correctly linking scores back to the attention-direction dynamics that preceded them.

Because effort is both an outcome of ADA and a potential pathway to increased performance, we do not treat effort as a standard control in the main regressions; instead, we formally model its mediating role.

### *Mediation Analysis*

To examine whether the effect of ADA on grant-period performance operates partly through increased communicative effort, we conducted a single-mediator causal analysis using team-round–level data. For all *Channel* × *Grant-period* observations belonging to a Grant round, we used the first factor score from the slope-based factor analysis as the measure of ADA (the treatment). Messaging effort was operationalized as the total number of messages sent in that channel during the same Grant period, which served as the mediator. Performance was measured along two dimensions: (i) the reviewer score assigned to each evaluated submission, with all scores retained when teams submitted multiple proposals in the same round; and (ii) the number of proposals submitted by the team in that round.

For both outcomes, we estimated the standard two-equation system:

$$Effort_i = a_0 + a_1 ADA_i + \varepsilon_i$$
$$Y_i = b_0 + b_1 Effort_i + c\ ADA_i + \xi_i$$

where $Y_i$ denotes either the score or the submission count for team-round $i$. The first equation tests whether higher ADA predicts greater communicative effort within the round. The second equation tests whether this effort, in turn, predicts performance, while simultaneously estimating the residual direct effect of ADA that does not operate through messaging.

From these equations, we computed the Average Causal Mediation Effect (ACME), which captures the portion of ADA's influence on performance that operates through messaging effort and is given by the product $a_1\ b_1$. The Average Direct Effect (ADE) corresponds to the coefficient $c$ and represents the effect of ADA not transmitted through the mediator. Their sum yields the Total Effect. Both equations were estimated using ordinary least squares and combined using the *mediate* algorithm (Imai et al.), which implements nonparametric causal mediation analysis. We used $1{,}000$ bootstrap replications to obtain empirical confidence intervals for ACME, ADE, and the Total Effect, providing inference robust to non-normal residuals and heteroskedasticity.

### *Robustness checks*

We conduct a series of robustness checks to evaluate whether our findings depend on modeling choices related to data coverage, temporal aggregation, and the dimensionality of the attention construct (see Tables S4–S6 and Figures S6–S7).

First, we re-estimate all models while including activity occurring prior to channel creation when computing Attention-Directing Ability. This extension is theoretically motivated by the fact that coordination, task allocation, and learning often begin well before formal project channels are created. To avoid introducing unrelated activity, we restrict this extension to actions performed by users who later participate in the focal channel, and we limit the observation window to the one-month period preceding channel creation. Empirically, this choice substantially increases the amount of information used to estimate attention dynamics, resulting in 141,754 actions aggregated into 669,659 temporal bins. Across both maximum likelihood and two-step Heckman specifications, the estimated effect of Attention-Directing Ability on submission likelihood remains positive and statistically significant (selection equation: $\beta \approx 1.07$–$1.16$), as does its effect on evaluation scores (outcome equation: $\beta \approx 2.00$–$2.41$; Table S3). The magnitude and significance of the selection correction terms ($\rho \approx 0.74$ in ML; *IMR ≈ 0.96* in the two-step model) further indicate that the core results are not driven by sample truncation.

Second, we assess sensitivity to the temporal resolution used to compute attention dynamics by increasing the bin size from 1 hour to 1 day. This test addresses the concern that fine-grained temporal aggregation may overfit short-term fluctuations in activity. Despite the coarser aggregation, Attention-Directing Ability continues to positively predict both submission and evaluation outcomes, with effect sizes comparable to the baseline specification (selection equation: $\beta \approx 1.16–1.18$; outcome equation: $\beta \approx 1.16–1.27$; Table S4). The direction and order of magnitude of the coefficients remain stable across estimation methods, indicating that the results do not hinge on a particular choice of temporal granularity.

Third, we examine whether the results depend on the latent structure of attention by replacing the single-factor Attention-Directing Ability measure with a three-factor solution derived from exploratory factor analysis. The EFA loadings (Figure S6a) reveal three distinct attention profiles. Across specifications, the first factor consistently captures the largest share of variance and exhibits the strongest association with both submission and evaluation outcomes, indicating that it represents the dominant attention direction signal. Importantly, mediation analyses show that indirect (ACME), direct (ADE), and total effects remain positive across all three factors, whereas direct effects (ADE) are smaller and less stable, with wider uncertainty intervals for the weaker factors (Figure S6b–c), supporting the robustness of the mediation results to alternative latent representations. The remaining factors capture qualitatively different attention architectures, such as early activation dynamics or external coordination signals, which may correspond to alternative modes of collective attention and coordination. We leave a systematic investigation of these distinct attention patterns and their potential functional roles to future work.

### ***Sensitivity to slop aggregation rule.***

We examined whether the main findings are sensitive to the aggregation rule used to construct ADA from the estimated random slopes, comparing the baseline positive-slope specification to net responsiveness and total interaction magnitude.

In the baseline specification, we average only the positive random slopes for each trigger action across all response actions, isolating upward mobilization dynamics. We compared this to two alternatives: the sum of all slopes regardless of sign, which captures net responsiveness, and the sum of absolute slope values, which captures total interaction magnitude without regard to direction.

Both alternative specifications yield factor loading vectors substantially correlated with the baseline ($r = 0.83$ for net responsiveness; $r = 0.81$ for total magnitude), with the core action types (first direct interactions, dyadic mentions, messages, replies, and reactions) loading in the same direction and rank order. Factor scores are more convergent for total magnitude ($r = 0.86$ with the baseline) than for net responsiveness ($r = 0.53$), reflecting the fact that signed averaging dilutes the common structure when suppressive slopes partially cancel mobilizing ones.

Predictive validity, however, diverges across specifications in theoretically informative ways. The net responsiveness operationalization shows no significant association with either submission or performance in the selection-corrected models, and mediation effects are near zero and non-significant for both outcomes. Total magnitude replicates the Heckman findings—ADA predicts submission likelihood (*$\beta \approx$ 1.64–1.67, $p < .001$*) and evaluation score conditional on submission (*$\beta \approx$ 1.52–1.96, $p < .001$* and *$p = .013$*)—but the mediation mechanism partially breaks down: the indirect effect through communication effort remains directionally consistent for submission (*average causal mediation effect = 0.54, $p = .076$*) but is non-significant for score (*average causal mediation effect = 0.18, $p = .75$*).

These results indicate that the positive-slope filter is an important design choice. Net responsiveness fails because signed cancellation—averaging mobilizing and suppressive dynamics—destroys the predictive signal. Total magnitude recovers the direct association between ADA and outcomes but loses the mediation for score, suggesting that absolute interaction intensity and directionally positive mobilization have distinct downstream consequences: the pathway through sustained collective effort that drives performance quality is specifically a property of upward responsiveness, not of total interaction magnitude. The baseline specification is thus the only one that jointly identifies both the performance effect and the mechanism through which it operates, consistent with the interpretation that what distinguishes high-ADA teams is not how reactive they are in general but how effectively their signals mobilize goal-directed engagement.

### *Temporal stability of Attention-Directing Ability*

We assessed the temporal coherence of Attention-Directing Ability across successive grant rounds as a test of construct validity, examining lag-1 and lag-2 autocorrelations, round-to-round Pearson and Spearman correlations, and the pattern of stability across the program. We examined the autocorrelation structure of ADA scores across successive grant rounds within teams. Lag-1 and lag-2 Pearson correlations across consecutive and skip-one grant periods are *$r = 0.32$ ($p < .001$, $N = 161$ pairs)* and *$r = 0.24$ ($p = .014$, $N = 106$ pairs)* respectively, confirming that a team's ADA in one round carries meaningful predictive information about its ADA in subsequent rounds.

Notably, the stability of ADA increases over the course of the program. Round-to-round Pearson correlations rise monotonically from *$r = 0.17$ (rounds 1–2, $p = .294$)* to *$r = 0.68$ (rounds 4–5, $p < .001$)*, and the long-range correlation between round 1 and round 5 ADA is $r = 0.71$ among the 22 teams observed across all five rounds. Spearman rank correlations between consecutive rounds range from *$\rho = 0.46$ to $\rho = 0.84$*, indicating that the relative ordering of teams by attention-directing ability is highly preserved as the program unfolds. This increasing stability is consistent with a consolidation process in which teams that sustain participation progressively develop and stabilize characteristic coordination styles. We note, however, that this pattern is also subject to potential survivor bias: only teams that remain active appear in later round pairs, and such teams may be more coordinated to begin with. The increasing stability estimates across rounds could therefore partly reflect the progressive selection of more coherent teams rather than within-team consolidation alone, and should be interpreted with this caveat in mind. Taken together, these analyses support the interpretation of ADA as a team-level capability with meaningful temporal coherence, while leaving open the question of whether the strengthening of that coherence over time reflects genuine learning and consolidation or the attrition of less coordinated teams from the sample.

## SUPPLEMENTARY FIGURES

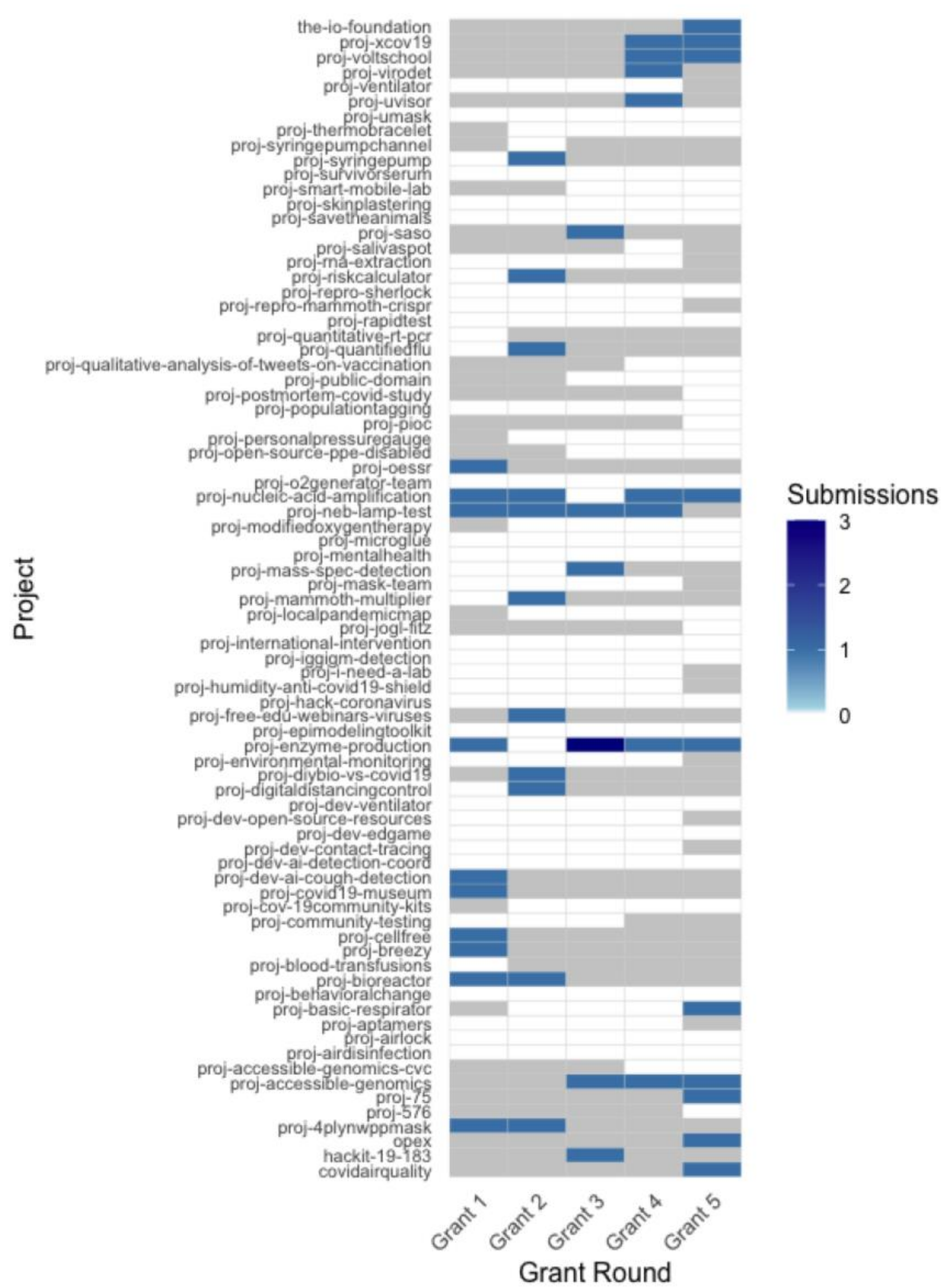

***Figure S1. Submission activity across projects and grant rounds.*** Heatmap showing all projects included in the OpenCovid19 dataset (rows) and the five grant rounds (columns). Cell color indicates the number of submissions made by each project in each round (0–3). Gray cells denote project–round combinations in which no channel existed or no activity was recorded. This figure documents the empirical structure of participation across rounds and the distribution of observed submission opportunities that underlie the selection and outcome analyses reported in the main text.

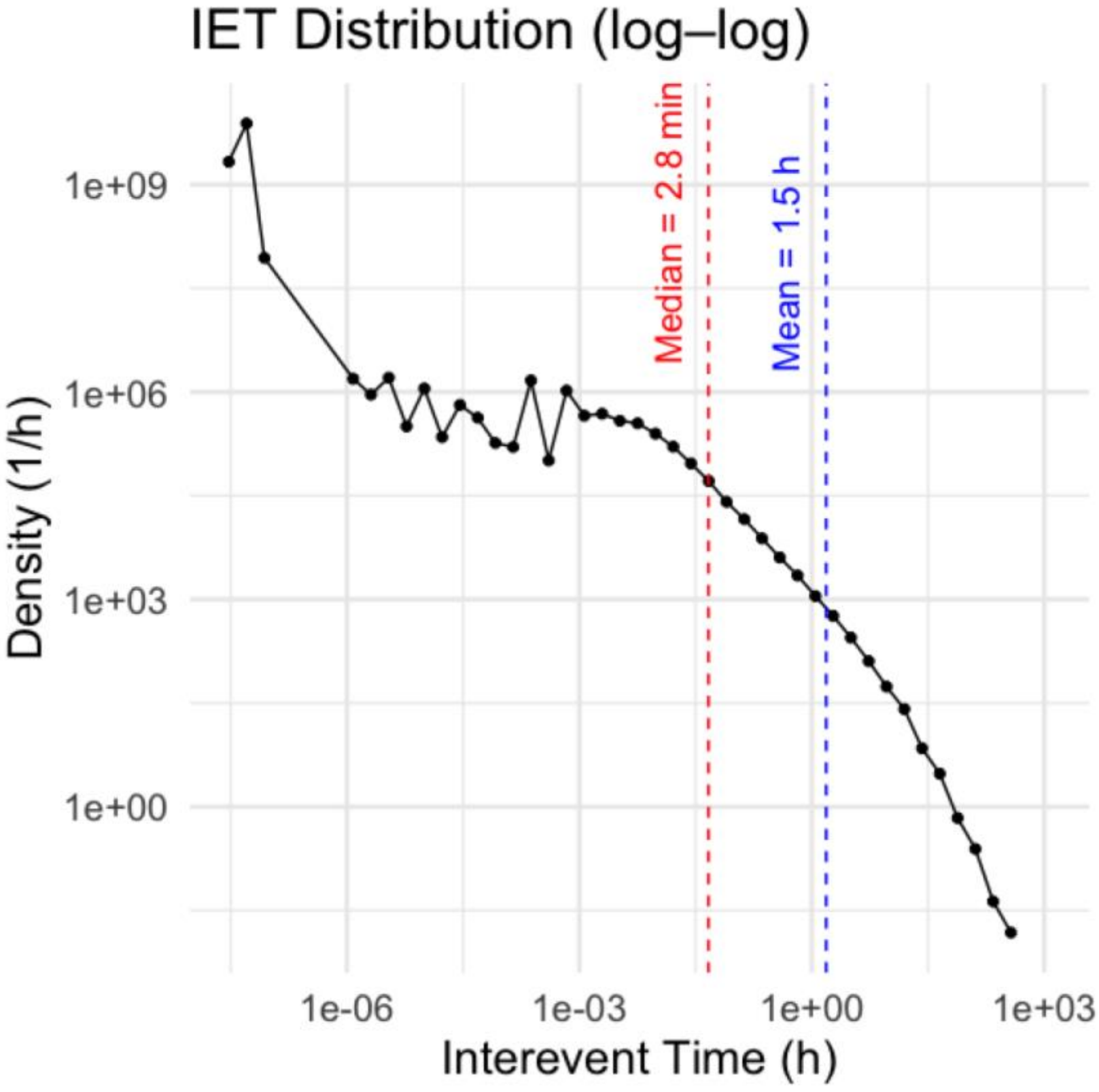

***Figure S2. Inter-event time distribution for Slack message activity (log–log scale).*** The distribution of inter-event times between consecutive messages is highly heavy-tailed, spanning over six orders of magnitude. The median inter-event time is approximately 2.8 minutes, while the mean is 1.5 hours (vertical dashed lines). This bursty temporal structure is characteristic of human communication and justifies aggregating activity into 1-hour windows when estimating lagged action–reaction effects: a 1-hour bin captures the full decay of short-term responses while smoothing over rare long gaps. The observed scale separation ensures that the modeled lag captures meaningful behavioral influence rather than noise from microbursts or sparse intervals.

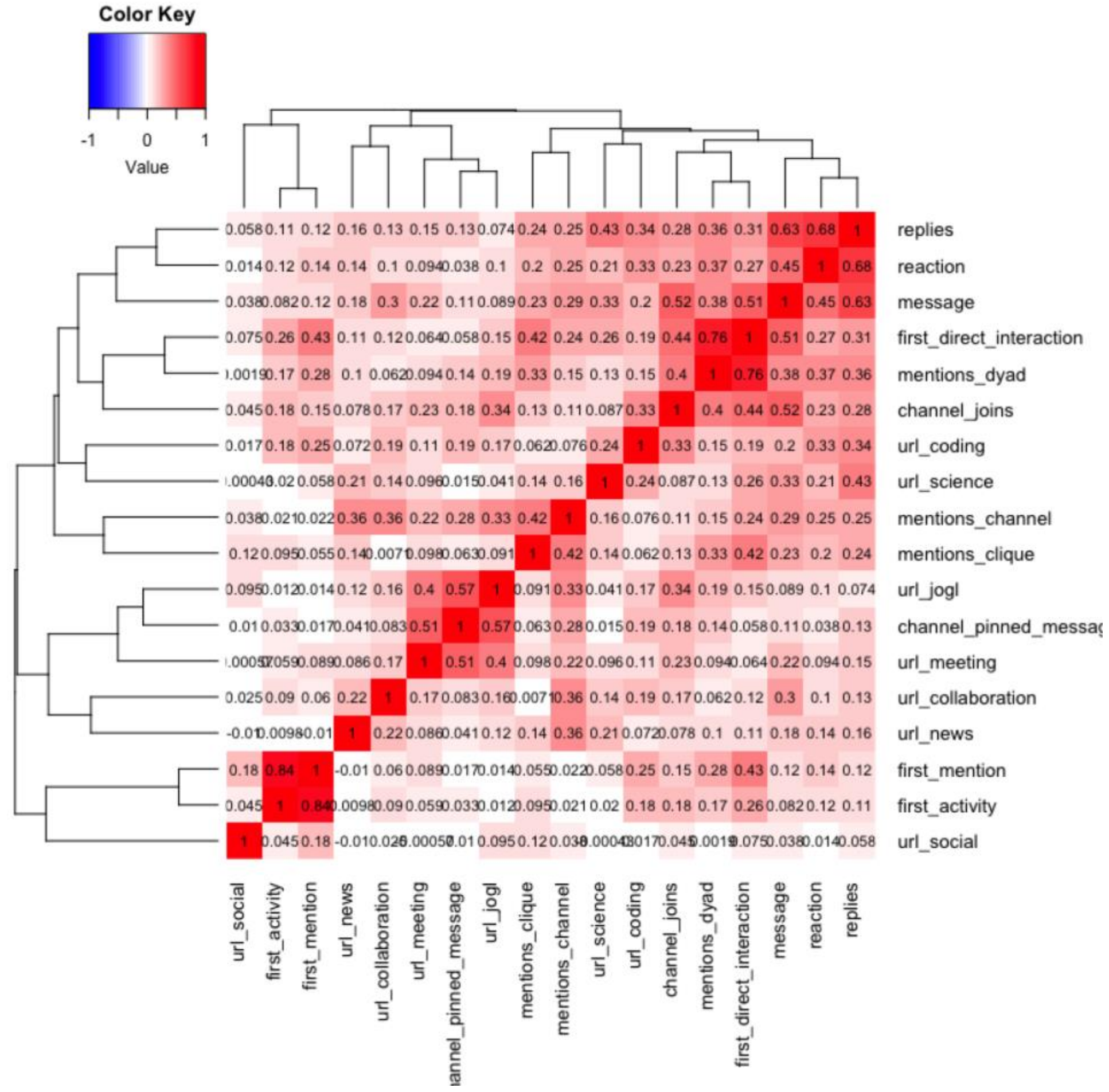

***Figure S3. Correlation structure of action-level responsiveness profiles.*** Heatmap of pairwise Pearson correlations among the 18 action-level responsiveness profiles, computed across all channel–grant-period observations, with hierarchical clustering applied to rows and columns. Red indicates positive correlations and blue indicates negative correlations. The strong block-diagonal structure reveals that channels displaying strong coupling for one type of behavioral response (e.g., conversational actions such as messages, replies, and mentions) tend to display strong couplings for related actions as well. This correlated structure provides the empirical basis for extracting a low-dimensional latent factor (ADA) from the channel-specific response profiles.

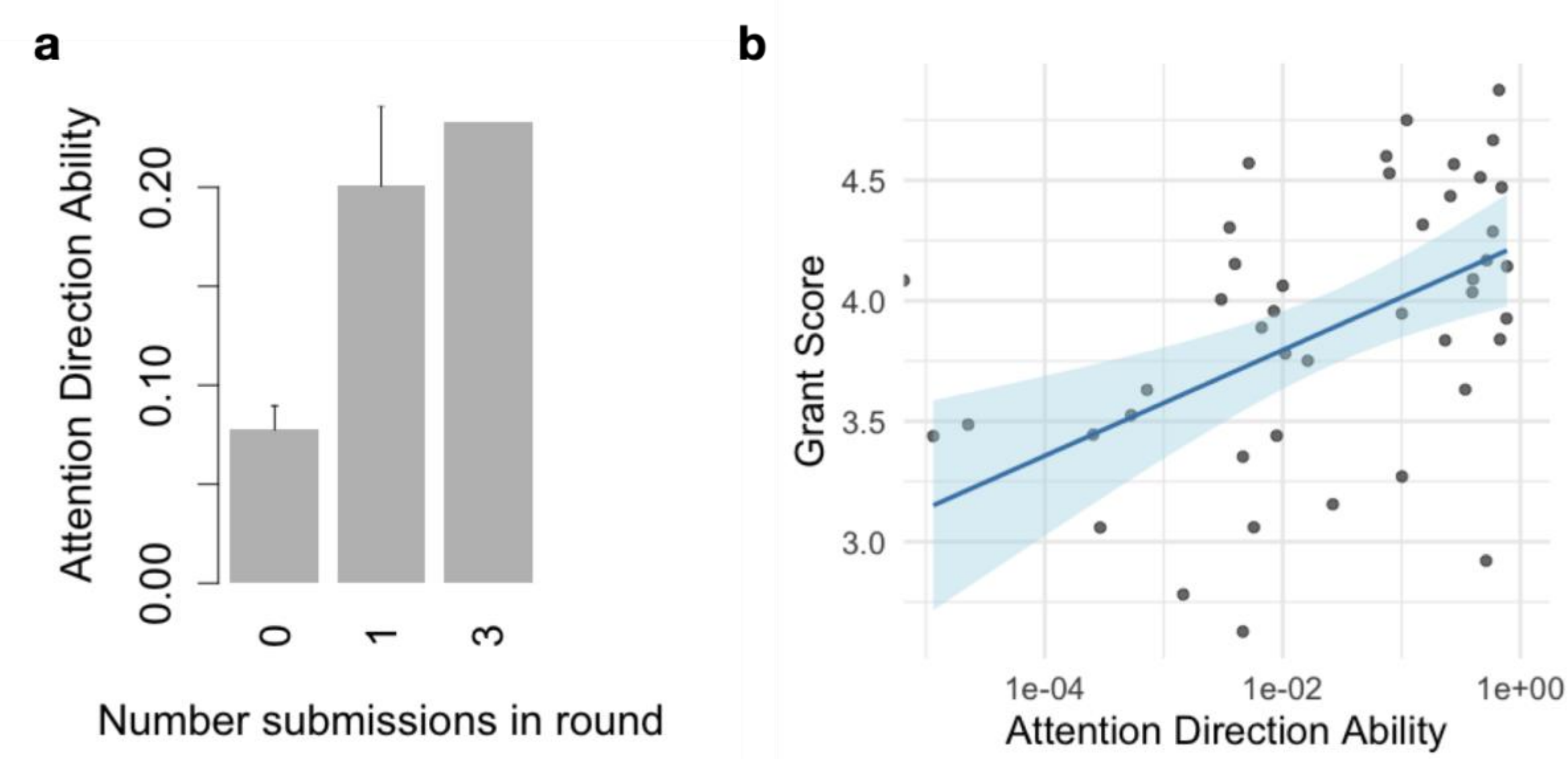

***Figure S4. Descriptive relationships between ADA, grant submission, and evaluation score.*** (a) Channels that submitted at least one proposal in a grant round exhibit substantially higher ADA than those that did not submit. Bars show mean ADA and standard errors across channels with 0, 1, or 3 submissions. (b) Among submitting teams (N = 44), grant evaluation scores increase monotonically with ADA. Points show individual channels; the line and shaded band show an OLS fit with 95% confidence intervals. These descriptive patterns illustrate the monotonic association between ADA and outcomes and complement the selection-corrected estimates in the main text. Inferential claims are based on the Heckman model and mediation analysis reported in Figures 2–3.

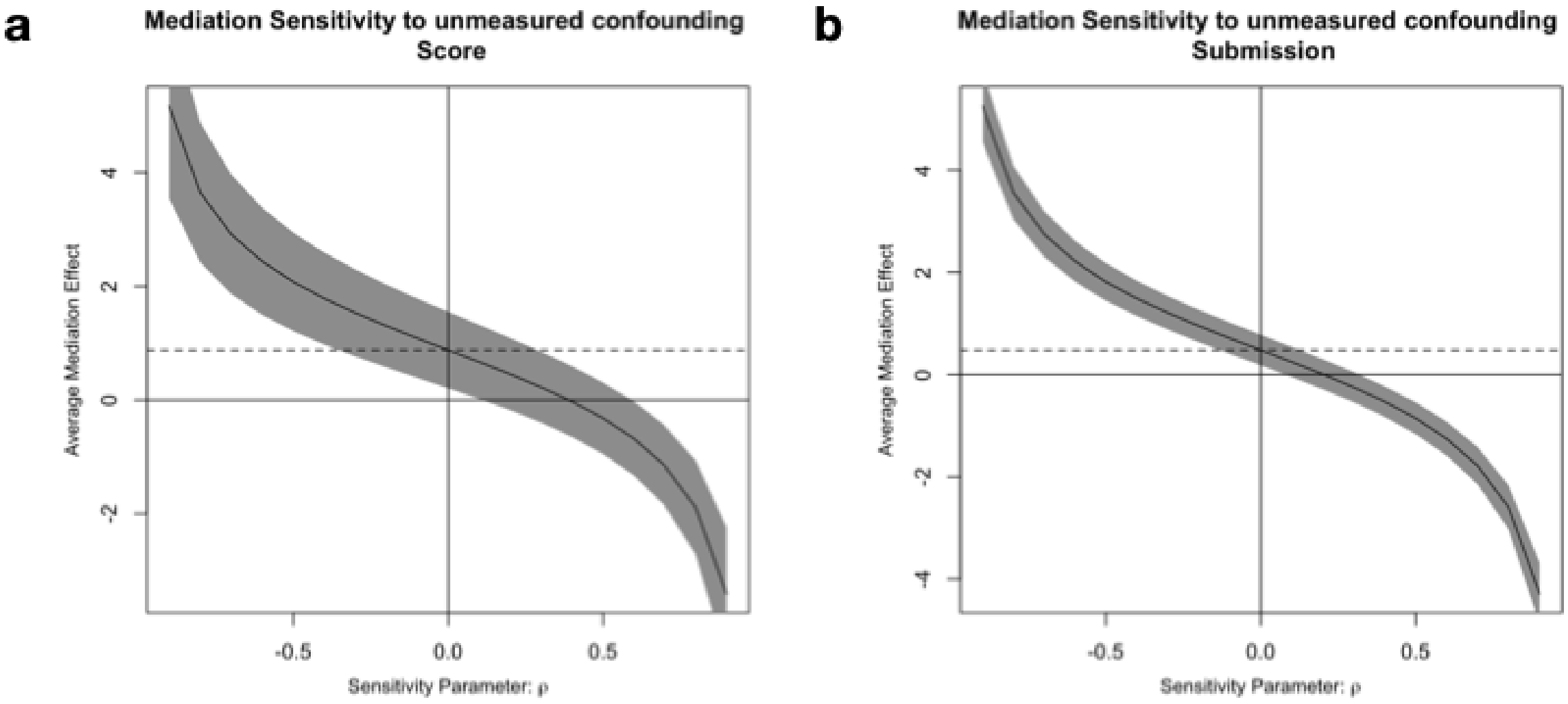

***Figure S5. Sensitivity of mediation effects to unmeasured confounding***. Each panel shows the estimated average causal mediation effect (ACME) as a function of the sensitivity parameter ρ, which represents the hypothetical correlation between unmeasured confounders of the mediator–outcome relationship. The solid line shows the point estimate and the shaded band the 95% confidence interval across 1,000 bootstrap replications. The horizontal solid line indicates ACME = 0; the dashed line indicates the observed ACME at $\rho = 0$. The left panel shows results for evaluation score: the ACME crosses zero at $\rho \approx 0.4$, indicating that a confounder would need to explain at least 16% of residual variance in both the mediator and outcome equations to nullify the mediated effect. The right panel shows results for submission: the zero-crossing occurs at $\rho \approx 0.2$. In both cases the ACME is positive and the confidence interval excludes zero for small values of ρ, supporting the robustness of the mediation findings to moderate unmeasured confounding.

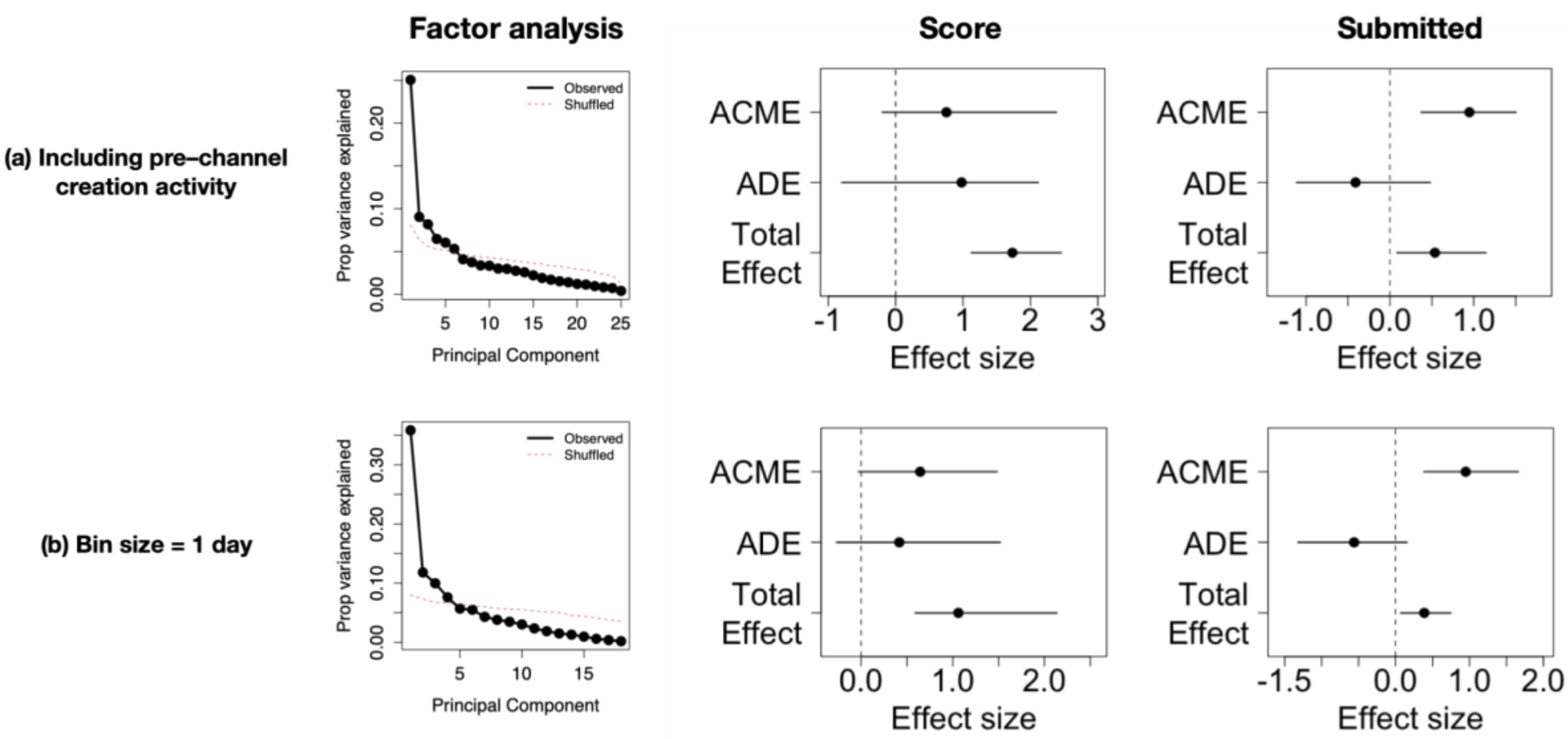

***Figure S6. Robustness of mediation effects to data window and temporal aggregation.*** Mediation effects of Attention-Directing Ability on evaluation score (left column) and submission decision (right column) under two alternative specifications. The top row shows results when including activity from the one-month period prior to channel creation (restricted to users of the focal channel). The bottom row shows results when attention dynamics are computed using 1-day temporal bins instead of 1-hour bins. Points denote estimates of indirect (ACME), direct (ADE), and total effects, with horizontal bars indicating 95% confidence intervals. The dashed vertical line indicates zero effect. Results remain consistent in sign and magnitude across specifications.

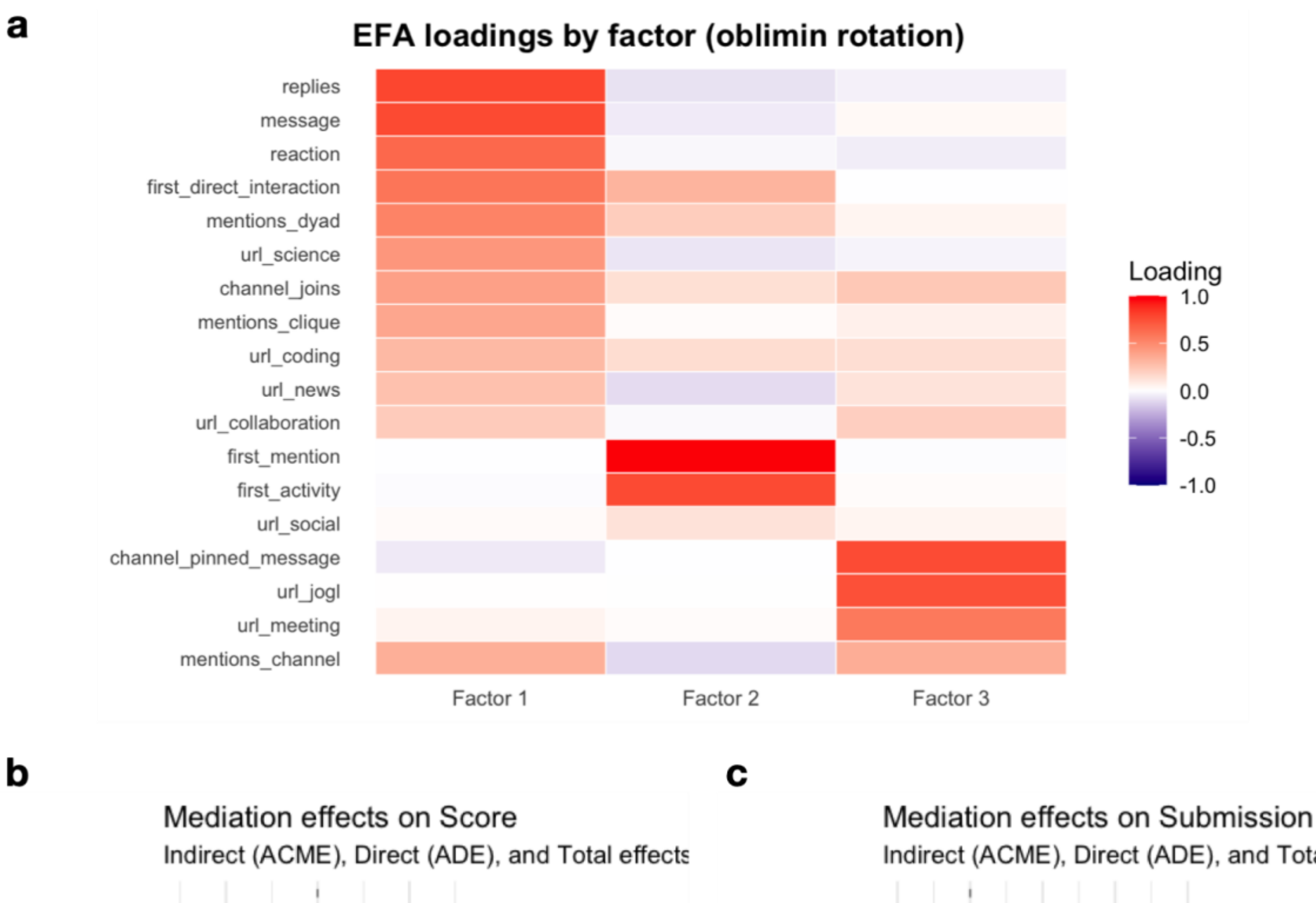

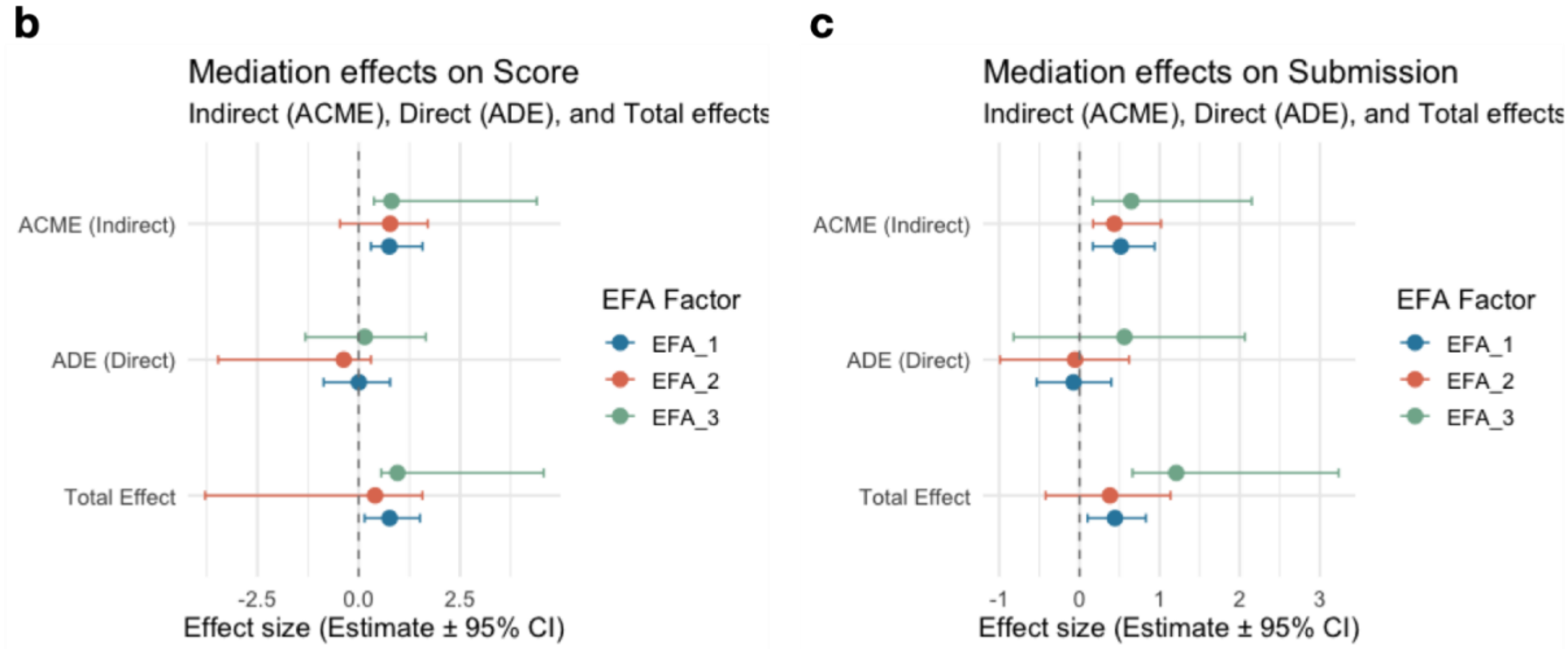

***Figure S7. Robustness of mediation effects across a three-factor attention structure.*** Panel (a) shows the exploratory factor analysis (EFA) loadings of attention-related activity features using oblimin rotation, revealing three distinct attention profiles. Panels (b) and (c) report mediation effects of Attention-Directing Ability on evaluation score and submission decision, respectively, for each factor. Points denote estimates of indirect (ACME), direct (ADE), and total effects, with horizontal bars indicating 95% confidence intervals; the dashed vertical line indicates zero effect. While the factors capture qualitatively different attention architectures, mediation results remain consistent in sign and magnitude across factors, indicating that the substantive conclusions do not depend on the dimensionality of the attention construct.

## SUPPLEMENTARY TABLES

***Table S1. Causal mediation analysis for evaluation score.***

| Effect | Estimate | CI_Lower | CI_Upper | p_value | Signif |
|---|---|---|---|---|---|
| ACME | 0.880 | 0.131 | 1.69 | 0.026 | * |
| ADE | 0.107 | -0.829 | 1.21 | 0.808 | |
| Total effect | 0.987 | 0.482 | 1.70 | <2e-16 | *** |
| Proportion mediated | 0.891 | 0.154 | 2.37 | 0.026 | * |

Notes: ACME = average causal mediation effect; ADE = average direct effect. Percentile bootstrap CI (1000 simulations). Sample size N = 44. Significance: *** $p < 0.001$, ** $p < 0.01$, * $p < 0.05$.

***Table S2. Causal mediation analysis for submission decision.***

| Effect | Estimate | CI_Lower | CI_Upper | p_value | Signif |
|---|---|---|---|---|---|
| ACME | 0.4817 | 0.0605 | 0.93 | 0.026 | * |
| ADE | 0.1420 | -0.3247 | 0.69 | 0.564 | |
| Total effect | 0.6237 | 0.2595 | 1.03 | <2e-16 | *** |
| Proportion mediated | 0.7724 | 0.0994 | 1.76 | 0.026 | * |

Notes: ACME = average causal mediation effect; ADE = average direct effect. Percentile bootstrap CI (1000 simulations). Sample size N = 240. Significance: *** $p < 0.001$, ** $p < 0.01$, * $p < 0.05$.

***Table S3. Example dialogues drawn from high-performing teams.*** Presents illustrative action–reaction sequences from high-performing teams (grant score > 4.5). We focus on interactions within the core interpersonal coordination network (i.e., dyadic mentions, first direct interactions, replies, and messages) where attention direction is most observable. The examples illustrate how attention-directing signals trigger downstream engagement, recruit expertise, and generate coordinated action.

| **Attention-directing signal (Action)** | **Interaction type** | **Downstream response (Reaction)** | **Coordination outcome** |
|---|---|---|---|
| *" if anyone has any access or ideas about a COVID-19 cough dataset…"* | Broad call for expertise (message) | *"<@UV90DGB4N> is really good with ReactJS app dev, I will invite him to this channel."* Followed by a new member joining and discussing ICU doctors, smartphone recording specifications, and data collection requirements. | Recruitment of expertise and advancement of technical problem solving. |
| *"I contacted hospital doctors at ICU… they need to see people behind the project so they can trust you and invest their time…"* | Resource mobilization and request for commitment (message) | *"Got it! No problem… I will allocate as much time as I can to building a prototype."* Additional members accepted invitations and connected external practitioners and researchers. | Increased member commitment and access to external stakeholders. |
| *"Anyone with design skills please request access… we need to act fast… If you want to jump to the React code, please coordinate with <@UV90DGB4N>."* | Collective call to action (channel mention) | Members volunteered skills, coordinated coding efforts, and organized responsibilities around design, machine learning, and application development. | Task allocation and accelerated collective progress. |
| *"Welcome <@U010KAJ3SLW>… we could benefit from* | First direct interaction / onboarding | The newcomer was directed to previous | Integration of new expertise into |

| *your experience in TensorFlow and time-series."* | | discussions and specific technical challenges requiring attention. | ongoing team work. |
|---|---|---|---|
| *"Can we use the data from [dataset]? Someone would ask for it?"* | Information-seeking signal | Team members shared additional papers, datasets, technical constraints, and implementation considerations. | Knowledge integration and refinement of project decisions. |

***Table S4. Sample-selection models including activity prior to channel creation (bin = 1 hour).***

| | **Maximum Likelihood estimation** | | **2-step Heckman / heckit estimation** | |
|---|---|---|---|---|
| **Variables** | **Coef.** | **S.E.** | **Coef.** | **S.E.** |
| **Selection equation:** | | | | |
| Intercept | -0.825*** | (0.140) | -0.867*** | (0.147) |
| Attention-Directing Ability | 1.072. | (0.552) | 1.156* | (0.557) |
| Days since creation | -0.002* | (0.001) | -0.002. | (0.001) |
| **Outcome equation:** | | | | |
| Intercept | 2.980*** | (0.432) | 2.214* | (1.116) |
| Attention-Directing Ability | 2.004*** | (0.510) | 2.407** | (0.869) |
| **Error terms:** | | | | |
| σ (sigma) | 0.612*** | (0.173) | 0.959 | (NA) |
| ρ (rho) | 0.739** | (0.252) | 1.002 | (NA) |
| Inverse Mills ratio | | | 0.960 | (0.724) |

Notes: Standard errors in parentheses. N = 240 (196 censored, 44 observed). σ denotes the outcome equation error standard deviation; ρ denotes the correlation between selection and outcome errors. ML log-likelihood = −137.920. *** $p < 0.001$, ** $p < 0.01$, * $p < 0.05$, . $p < 0.1$.

***Table S5. Sample-selection models with daily temporal aggregation (bin = 1 day)***

| | **Maximum Likelihood estimation** | | **2-step Heckman / heckit estimation** | |
|---|---|---|---|---|

| Variables | Coef. | S.E. | Coef. | S.E. |
|---|---|---|---|---|
| **Selection equation:** | | | | |
| Intercept | -0.843*** | (0.144) | -0.855*** | (0.147) |
| Attention Direction Ability | 1.159* | (0.484) | 1.182* | (0.489) |
| Days since creation | -0.002* | (0.001) | -0.002. | (0.001) |
| **Outcome equation:** | | | | |
| Intercept | 3.423*** | (0.526) | 3.189*** | (0.838) |
| Attention Direction Ability | 1.158** | (0.432) | 1.266* | (0.551) |
| **Error terms:** | | | | |
| σ (sigma) | 0.523*** | (0.142) | 0.598 | (NA) |
| ρ (rho) | 0.484 | (0.538) | 0.691 | (NA) |
| Inverse Mills ratio | | | 0.410 | (0.554) |

Notes: Standard errors in parentheses. N = 240 (196 censored, 44 observed). σ denotes the outcome equation error standard deviation; ρ denotes the correlation between selection and outcome errors. ML log-likelihood = −129.421 *** $p < 0.001$, ** $p < 0.01$, * $p < 0.05$, . $p < 0.1$.

***Table S6. Sample-selection models using a three-factor attention structure***

| | Maximum Likelihood estimation | | 2-step Heckman / heckit estimation | |
|---|---|---|---|---|
| **Variables** | **Coef.** | **S.E.** | **Coef.** | **S.E.** |
| **Selection equation:** | | | | |
| Intercept | -0.853*** | (0.138) | -0.882*** | (0.143) |
| Attention Direction Ability | 1.173** | (0.450) | 1.194** | (0.456) |
| Days since creation | -0.002* | (0.001) | -0.002. | (0.001) |
| **Outcome equation:** | | | | |
| Intercept | 3.094*** | (0.484) | 2.489* | (1.233) |
| Attention Direction Ability | 1.100* | (0.443) | 1.421. | (0.747) |
| **Error terms:** | | | | |
| σ (sigma) | 0.650*** | (0.182) | 0.898 | (NA) |
| ρ (rho) | 0.682* | (0.303) | 0.936 | (NA) |
| Inverse Mills ratio | | | 0.841 | (0.699) |

Notes: Standard errors in parentheses. N = 240 (196 censored, 44 observed). $\sigma$ denotes the outcome equation error standard deviation; $\rho$ denotes the correlation between selection and outcome errors. ML log-likelihood = −141.79 *** $p < 0.001$, ** $p < 0.01$, * $p < 0.05$, . $p < 0.1$.